\documentclass[runningheads]{llncs}
\usepackage[T1]{fontenc}
\usepackage{graphicx}
\usepackage[dvipsnames]{xcolor}

\usepackage{todonotes} 
\usepackage[normalem]{ulem}

\usepackage{amsmath}
\usepackage{amssymb}
\usepackage{mathtools}
\usepackage{microtype}
\usepackage{xspace}
\usepackage{semantic}
\usepackage{tikz}
\usetikzlibrary{arrows.meta, fit, calc}
\usepackage{listings}
\usepackage{lstautogobble}
\usepackage{needspace}
\usepackage{textgreek}
\usepackage{array}
\usepackage{booktabs}
\usepackage{float}      

\usepackage[numbers,sort&compress]{natbib}

\usepackage[
  colorlinks=true,
  linkcolor=black,
  citecolor=black,
  urlcolor=blue,
  bookmarksdepth=2
]{hyperref}
\usepackage{orcidlink}   
\renewcommand{\orcidID}[1]{\unskip\,\orcidlink{#1}}

\newcommand{\LAlpha}{\texorpdfstring{\ensuremath{\lambda^{\alpha}_{\mathtt{Ref}}}\xspace}{LambdaAlphaRef}}
\newcommand{\LCtx}{\texorpdfstring{\ensuremath{\lambda^{\mathtt{Ctx}}_{\mathtt{Ref}}}\xspace}{LambdaCtxRef}}
\newcommand{\compile}{\rightsquigarrow}
\newcommand{\LForallBox}{\ensuremath{\lambda_{\forall[\,]}}\xspace}   

\usepackage[autostyle=true]{csquotes}
\newcommand{\CodeQuoteOpen}{\mbox{\ttfamily\textasciigrave\kern-0.2em\textasciigrave}}
\newcommand{\CodeQuoteClose}{\mbox{\ttfamily\textquotesingle\kern-0.2em\textquotesingle}}
\newcommand{\CodeQuote}[1]{\CodeQuoteOpen#1\CodeQuoteClose}
\newcommand{\CodeInline}{\lstinline[style=examplecode,breaklines=false]}  

\newcommand{\CodeListingLambda}{\mbox{\normalfont\bfseries\textlambda}}
\newcommand{\CodeListingBigLambda}{\mbox{\normalfont\bfseries\textLambda}}
\newcommand{\CodeListingFun}{\mbox{\normalfont\textrightarrow}}
\newcommand{\CodeListingDoubleFun}{\mbox{\normalfont$\Rightarrow$}}
\newcommand{\CodeListingForall}{\mbox{\normalfont$\forall$}}
\newcommand{\CodeListingBullet}{\mbox{\normalfont\bfseries\textbullet}}
\newcommand{\CodeListingTurnstile}{\mbox{\normalfont$\vdash$}}
\newcommand{\CodeListingEmptyCtx}{\mbox{\normalfont$\emptyset$}}
\newcommand{\CodeListingSubstFun}{\mbox{\normalfont$\Rrightarrow$}}   
\newcommand{\CodeListingCompose}{\mbox{\normalfont$\circ$}}           
\newcommand{\EC}[2]{\ensuremath{\mathord{\text{\normalfont\bfseries\textcolor{#1}{#2}}}}}  
\newcommand{\CodeListingEC}[2]{\mbox{\normalfont\bfseries\textcolor{#1}{#2}}}                
\newcommand{\ECeps}{\EC{Turquoise}{\textepsilon}}
\newcommand{\ECalpha}{\EC{Plum}{\textalpha}}
\newcommand{\ECbeta}{\EC{MidnightBlue}{\textbeta}}
\newcommand{\ECgamma}{\EC{PineGreen}{\textgamma}}

\newcommand{\UNIT}{\texttt{Unit}}
\newcommand{\REFTY}[1]{\texttt{Ref}\; #1}
\newcommand{\SPLICE}[1]{{\sim}#1}
\newcommand{\REF}[1]{\texttt{ref}\; #1}
\colorlet{CtxEps}{Turquoise!35}
\colorlet{CtxAlpha}{Plum!20}
\colorlet{CtxBeta}{MidnightBlue!20}
\colorlet{CtxGamma}{PineGreen!20}
\newcommand{\CodeListingBadge}[1]{%
  \hfill
  \CodeListingSmash{\tikz[baseline=(badge.base)]%
    \node[
      fill=Gray!20,
      draw=Black,
      rounded corners=1pt,
      inner xsep=3pt,
      inner ysep=1pt,
      font=\scriptsize\ttfamily\color{Black}
    ] (badge) {#1};}%
}
\colorlet{CodeListingBackground}{Goldenrod!3!White}

\newcommand{\RuleTag}[1]{\textit{\textcolor{Black!60}{\scriptsize(#1)}}}
\makeatletter
\newcommand{\tightlist}{\def\@listi{\leftmargin\leftmargini
  \parsep \z@ \topsep 2\p@ \@plus1\p@ \@minus1\p@ \itemsep 1\p@}}  
\makeatother

\newcommand{\CodeListingMark}[2]{%
  \tikz[remember picture,baseline=(#1.base)]%
    \node[inner sep=0pt,outer sep=0pt] (#1) {#2};%
}
\newcommand{\CodeListingSmash}[1]{\raisebox{0pt}[0pt][0pt]{#1}}
\newcommand{\CodeListingVarBoxRaw}[2]{{\setlength{\fboxsep}{1.5pt}\colorbox{#1}{\ttfamily\small #2}}}
\newcommand{\CodeListingVarBox}[2]{\CodeListingSmash{\CodeListingVarBoxRaw{#1}{#2}}}
\newtoks\CodeListingDeferredFrames
\newcounter{CodeListingFrameCount}
\newcommand{\CodeListingFrameBox}[2]{%
  \stepcounter{CodeListingFrameCount}%
  \CodeListingSmash{\CodeListingMark{CodeListingFrame\theCodeListingFrameCount}{\ttfamily\small #2}}%
  \begingroup\edef\CodeListingTmp{\global\CodeListingDeferredFrames{\the\CodeListingDeferredFrames
      \noexpand\CodeListingDrawFrame{#1}{CodeListingFrame\theCodeListingFrameCount}}}%
  \CodeListingTmp\endgroup}
\newcommand{\CodeListingDrawFrame}[2]{\tikz[remember picture,overlay] \node[draw=#1,line width=0.6pt,inner sep=1pt,fit=(#2)] {};}
\newcommand{\CodeListingFlushFrames}{%
  \edef\CodeListingTmp{\the\CodeListingDeferredFrames}%
  \ifx\CodeListingTmp\empty\else
    \dimen255=\prevdepth\nointerlineskip
    \hbox{\the\CodeListingDeferredFrames}\prevdepth=\dimen255
    \global\CodeListingDeferredFrames{}%
  \fi}
\AddToHook{env/lstlisting/after}{\CodeListingFlushFrames}
\newcommand{\CodeListingUnderline}[2]{\CodeListingSmash{\setbox0=\hbox{\ttfamily\small #2}%
  \rlap{\color{#1}\rule[-3.1pt]{\wd0}{0.8pt}}\copy0}}
\newcommand{\CodeListingVarMark}[3]{\CodeListingSmash{\CodeListingMark{#2}{\CodeListingVarBoxRaw{#1}{#3}}}}
\newcommand{\CodeHighlight}[2]{\CodeListingVarBoxRaw{#1}{#2}}
\tikzset{
  binding line/.style={line width=1pt, shorten <=0.3pt, shorten >=0.3pt},
  binding arrow/.style={binding line, -{Stealth[length=2mm]}},
}
\newcommand{\CodeListingArrow}[4]{%
  \tikz[remember picture,overlay] \draw[binding arrow, #1] (#2) to[bend #4] (#3);%
}
\newcommand{\CodeListingArrowVia}[6]{%
  \tikz[remember picture,overlay]
    \draw[binding arrow, #1] (#2) to[bend #4] coordinate[pos=#5] (#6) (#3);%
}
\newcommand{\CodeListingLine}[4]{%
  \tikz[remember picture,overlay] \draw[binding line, #1] (#2) to[bend #4] (#3);%
}

\lstdefinestyle{tt}{
  basicstyle=\ttfamily\small,
  keywordstyle=\bfseries,
  morekeywords={let, in, ref, if, then, else, run, lambda, for, to, do, done, rec},
  moredelim = [s][\color{Gray}]{/*}{*/},
  literate={``}{{\CodeQuoteOpen}}2
           {''}{{\CodeQuoteClose}}2
           {->}{{\CodeListingFun}}2
           {=>}{{\CodeListingDoubleFun}}2
           {∀}{{\CodeListingForall}}1
           {•}{{\CodeListingBullet}}1
           {λ}{{\CodeListingLambda}}1
           {Λ}{{\CodeListingBigLambda}}1
           {⊢}{{\CodeListingTurnstile}}1
           {∅}{{\CodeListingEmptyCtx}}1
           {⇛}{{\CodeListingSubstFun}}1
           {∘}{{\CodeListingCompose}}1
           {ε}{{\CodeListingEC{Turquoise}{\textepsilon}}}1
           {α}{{\CodeListingEC{Plum}{\textalpha}}}1
           {β}{{\CodeListingEC{MidnightBlue}{\textbeta}}}1
           {γ}{{\CodeListingEC{PineGreen}{\textgamma}}}1
           {ξ}{{\CodeListingEC{Tan}{\textxi}}}1,
  autogobble,
  escapechar=|
}

\lstdefinestyle{examplecode}{
  style=tt,
  columns=fullflexible,
  keepspaces=true,
  showstringspaces=false,
  breaklines=true,
  numbers=left,
  numberstyle=\scriptsize\color{Gray},
  numbersep=0.5em,
  frame=single,
  framerule=0.3pt,
  rulecolor=\color{Gray},
  backgroundcolor=\color{CodeListingBackground},
  xleftmargin=0.75em,
  xrightmargin=0.75em,
  framexleftmargin=0.8em,
  framexrightmargin=0.5em,
  aboveskip=0.75\baselineskip,
  belowskip=0.75\baselineskip
}

\newcommand{\RuleHighlight}[2][ProcessBlue!15]{{\setlength{\fboxsep}{1.5pt}\colorbox{#1}{$#2$}}}
\newcommand{\RuleTagHighlight}[2][ProcessBlue!15]{{\setlength{\fboxsep}{1.5pt}\colorbox{#1}{\RuleTag{#2}}}}

\newcommand{\Fbox}[1]{\fcolorbox{Black}{CodeListingBackground}{#1}}  

\newcommand{\Sem}[1]{\ensuremath{\mathopen{[\![}#1\mathclose{]\!]}}}            
\newcommand{\Erase}[1]{\ensuremath{\lvert #1 \rvert}}                           
\newcommand{\Reify}[1]{\ensuremath{\mathsf{reify}(#1)}}                          
\newcommand{\ReifyEnv}[2]{\ensuremath{\mathsf{reify}_{#1}(#2)}}                  
\newcommand{\Int}{\ensuremath{\mathsf{Int}}}
\newcommand{\QuoteOpen}{\CodeQuoteOpen}
\newcommand{\QuoteClose}[1][]{\CodeQuoteClose#1}
\newcommand{\Quote}[2]{\ensuremath{\QuoteOpen#1\QuoteClose[#2]}}         
\newcommand{\CQuote}[2]{\ensuremath{\QuoteOpen#1 \vdash #2\QuoteClose}}  

\newcommand{\SubstTy}[2]{\ensuremath{#1 \Rrightarrow #2}}        
\newcommand{\sub}{\mathrel{<:}}                                  
\newcommand{\simrel}{\mathrel{\thicksim}}                          
\newcommand{\coerceval}{\mathrel{\triangleright}}                
\newcommand{\IdSub}[1]{\ensuremath{\mathtt{id}_{#1}}}                  
\newcommand{\WeakenSub}[2]{\ensuremath{\mathtt{weaken}_{#1}(#2)}}      
\newcommand{\Proxy}[3]{\ensuremath{\mathtt{proxy}\;#1\;#2\;#3}}        
\newcommand{\Unit}{\ensuremath{\mathsf{Unit}}}
\newcommand{\unit}{\ensuremath{()}}

\begin{document}
\title{Compiling Environment Classifiers to \\ Contextual Modal Types}
%
%
\author{Tianyu~Chen\inst{1}\orcidID{0009-0002-3279-5971} \and
Darshal~Shetty\inst{1}\orcidID{0000-0001-8365-5763} \and
Jeremy~G.~Siek\inst{1}\orcidID{0000-0002-9894-4856} \\
Chao-Hong~Chen\inst{2}\orcidID{0000-0002-9314-8105} \and
Weixi~Ma\inst{2}\orcidID{0009-0003-1642-9880} \and
Arnaud~Venet\inst{2}\orcidID{0009-0008-3370-2021} \and
Rocky~Liu\inst{2}}
\authorrunning{T. Chen et al.}
%
\institute{Indiana University, Bloomington, IN, USA\\
\email{\{chen512,dcshetty,jsiek\}@iu.edu} \and
Meta, Menlo Park, CA, USA\\
\email{\{chaohong,mavc,ajv,rockyliu4\}@meta.com}}
\maketitle              

\begin{abstract}

Type systems for metaprogramming take one of two approaches to
safeguarding the scopes of variables in generated code. They either
use Environment Classifiers (EC) to identify scopes by name (Taha and
Nielsen, 2003), or they use Contextual Modal Type Theory (CMTT) to
match scopes by structure (Nanevski, Pfenning, and Pientka, 2008).
In this paper, we connect these two approaches by compiling a
calculus with polymorphic environment classifiers, subtyping, and
mutable references to a CMTT-based calculus with polymorphic contexts,
mutable references, and first-class substitutions. The key idea is to
map each environment classifier to a context. We prove that the
compilation preserves types and that compilation preserves the
behavior of metaprograms. Both proofs are mechanized in Agda. The
compilation shows that CMTT is at least as expressive as environment
classifiers. We conjecture that the converse compilation is difficult,
because environment classifiers impose a stricter regime on the
metaprogram.

\end{abstract}

\section{Introduction}
\label{sec:introduction}

Metaprogramming is the paradigm in which a program in a metalanguage generates a
program in a code language. In this paper, we study the relationship between two
typing disciplines for metaprogramming with open code, that is, code fragments
with free variables: environment classifiers~\cite{Taha:2003qo}, which name the
scope in which the code is meaningful, and contextual modal
types~\cite{Nanevski:2008aa}, which spell out that scope as a typing context.
Both guarantee that generated programs are well typed and well scoped, but they
have been developed largely independently. We show that every scope named by a
classifier can be translated into a context, define a compiler from a calculus
with environment classifiers to one with contextual modal types, and prove that
compilation preserves types and that source and target programs generate the
same code.

In the tradition of MetaML~\cite{Taha:1997ye,taha2000metaml}, the type system of
the metalanguage tracks the type of quoted code, so that every generated program
is well typed. The difficulty lies with open code: the free variables of a
fragment are bound by binders of the code language outside the fragment, so the
type system must also record the scope in which the fragment is meaningful.
Otherwise the fragment may leave that scope, through \texttt{run} or through a
mutable reference, and the generated program ends up with unbound variables, a
failure known as \emph{scope
extrusion}~\cite{kiselyov2014design,Kiselyov:2016aa,lee2026handling}.

\begin{figure}[t]
  \centering
  \begin{tikzpicture}[
      x=1cm, y=1cm,
      line/.style={line width=5pt},
      branch/.style={line, -{Triangle[length=3mm, width=5mm]}, rounded corners=3pt},
      pub/.style={font=\scriptsize, align=center, inner sep=2pt},
      header/.style={font=\small\bfseries, inner sep=2pt},
      calculus/.style={font=\normalsize, inner sep=3pt},
    ]
    \node[calculus, NavyBlue]    (lalpha) at (0,   -2.3) {\LAlpha};
    \node[calculus, BurntOrange] (lctx)   at (2.0, -2.3) {\LCtx};
    \draw[-{Stealth[length=2.5mm]}, line width=1.5pt]
      (lalpha) -- node[below, font=\footnotesize] (compile) {compile} (lctx);
    \node[draw=Red, dotted, line width=1pt, rounded corners=3pt,
          inner xsep=10pt, inner ysep=9pt, fit=(lalpha) (lctx) (compile)] (box) {};
    \node[anchor=east, font=\small\bfseries, Red] at ([xshift=-3pt]box.west) {\textit{This paper}};

    \coordinate (root)  at (1.0, 3.9);
    \coordinate (split) at (1.0, 3.3);
    \draw[line, gray!50] (root) -- (split);
    \draw[branch, NavyBlue!60]    (split) -| ([yshift=2pt]lalpha.north);
    \draw[branch, BurntOrange!60] (split) -| ([yshift=2pt]lctx.north);

    \node[header, NavyBlue,    anchor=east] at (-0.3, 3.62) {Environment Classifiers};
    \node[header, BurntOrange, anchor=west] at ( 2.3, 3.62) {Contextual Modal Types};

    \node[pub, anchor=east] at (0.7, 4.75) {%
      \textit{A temporal-logic approach to}\\
      \textit{binding-time analysis.}\\
      Davies~
      1996~\cite{davies1996temporal}\\
      ($\lambda^{\bigcirc}$: open code, but no \texttt{run})};
    \node[pub, anchor=east] at (-0.3, 2.4) {%
      \textit{Environment classifiers.}\\
      Taha and Nielsen\\
      2003~\cite{Taha:2003qo}\\
      ($\lambda^{\alpha}$: open code \textit{and} \texttt{run})};
    \node[pub, anchor=east] at (-0.3, 0.95) {%
      \textit{Refined environment classifiers.}\\
      Kiselyov, Kameyama, and Sudo\\
      2016~\cite{Kiselyov:2016aa}\\
      (\texttt{<NJ>}: mutable references)};
    \node[pub, anchor=east] at (-0.3, -0.75) {%
      \textit{Refined$^2$ environment classifiers.}\\
      Murase and Igarashi\\
      2026~\cite{murase2026refined}\\
      (R$^2$EC: classifier polymorphism)};

    \node[pub, anchor=west] at (1.3, 4.75) {%
      \textit{A modal analysis of}\\
      \textit{staged computation.}\\
      Davies and Pfenning~
      1996~\cite{Davies:1996gx}\\
      ($\lambda^{\Box}$: \texttt{run}, but only closed code)};
    \node[pub, anchor=west] at (2.3, 2.4) {%
      \textit{Contextual modal type theory.}\\
      Nanevski, Pfenning, and Pientka\\
      2008~\cite{Nanevski:2008aa}};
    \node[pub, anchor=west] at (2.3, 0.95) {%
      \textit{Contextual modal type theory}\\
      \textit{with polymorphic contexts.}\\
      Murase, Nishiwaki, and Igarashi\\
      2023~\cite{murase2023contextual}\\
      (\LForallBox: context polymorphism)};
    \node[pub, anchor=west] at (2.3, -0.75) {%
      \textit{Contextual MetaML.}\\
      Yin, Murawski, and Ong\\
      2026~\cite{yin2026contextual}\\
      (mutable references)};
  \end{tikzpicture}
  \caption{Two approaches to the scopes of generated code: environment
    classifiers (left) and contextual modal types (right). Both lines descend
    from the modal and temporal analyses of staged computation and are connected
    through compilation in this paper.}
  \label{fig:prior-work-two-lines}
\end{figure}
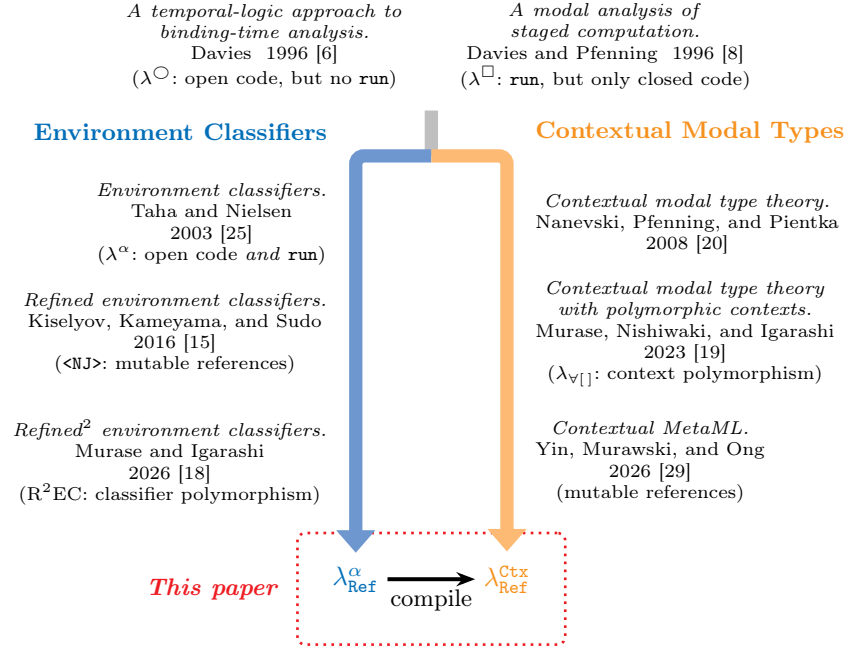


Two lines of research (Figure~\ref{fig:prior-work-two-lines}) have
developed type systems that track open code fragments and keep
generated code well scoped in
the presence of \texttt{run} or mutable references.  Both lines
descend from the Curry--Howard-style analyses of staged computation.
The calculus $\lambda^{\Box}$
of~\citet{Davies:1996gx,Davies:2001op} is based on the necessity
operator $\Box$ of the intuitionistic modal logic S4: a term of type
$\Box A$ is closed code of type $A$, which can safely be evaluated,
but the calculus manipulates closed code only. On the other hand, the
calculus $\lambda^{\bigcirc}$
of~\citet{davies1996temporal,Davies:2017aa} is based on the ``next''
operator $\bigcirc$ of linear-time temporal logic: a term of type
$\bigcirc A$ is code that may contain free variables. Thus
$\lambda^{\bigcirc}$ enables the manipulation of open code, which
makes it a foundation for the partial evaluation of functional
languages~\citep{Davies:2017aa}. However, $\lambda^{\bigcirc}$ does
not include \texttt{run} because its type system would not be able to
prevent scope extrusion in that
case~\citep{Davies:2017aa,Taha:2003qo}.

Practical metaprogramming needs both the manipulation of open code and
\texttt{run}. Environment classifiers start from $\lambda^{\bigcirc}$
and add \texttt{run}; contextual modal types start from
$\lambda^{\Box}$ and add support for open code.

\textbf{Environment classifiers.}
The research on environment classifiers, the left arrow of
Figure~\ref{fig:prior-work-two-lines}, grows out of $\lambda^{\bigcirc}$ and
MetaML, the latter of which combines open code with
\texttt{run}~\cite{Taha:1997ye,moggi1999idealized,taha2000metaml}. As an early
solution, \citet{calcagno2000closed,Calcagno:2003aa} propose closed types,
which single out the code types whose values are known to be closed.
They restrict \texttt{run} to code of closed type, and similarly
for constructing mutable references to code.
\citet{Taha:2003qo}
then introduce \emph{Environment Classifiers} (EC). An environment classifier is
a name of a scope, and a code type $\langle T \rangle^\alpha$ records the
classifier $\alpha$ of the scope in which its code, possibly open, is
meaningful. The type system checks by name that a code fragment only mentions
variables, and only splices code, of the same classifier. It permits
\texttt{run} only on code whose classifier is universally quantified, which
guarantees that the code is closed. Early versions of MetaOCaml adopted this
discipline~\cite{Calcagno:2003vn,Taha:2004ov}, and \citet{tsukada2010logical}
give it a logical foundation through a Curry--Howard correspondence with a
multi-modal logic.

\textbf{Refined environment classifiers.}
Mutable references reintroduce the scope extrusion problem, because
open code can be smuggled out of its scope by storing it in a
reference.  \citet*{Kiselyov:2016aa} rule out such scope extrusions
statically with \emph{refined} environment classifiers, which mark
every code-level binder with its own classifier, ordered by the
nesting of the binders, so that storing code in a reference cell with
an incompatible classifier is a type error.  \citet{isoda2024type} add
algebraic effects and handlers to refined environment classifiers, and
\citet{murase2026refined} extend them to multi-level staging with
\texttt{run}, cross-stage persistence, and polymorphism over
classifiers.

\textbf{Contextual modal types.}  The research on contextual modal
types, the right arrow of Figure~\ref{fig:prior-work-two-lines}, grows
out of $\lambda^{\Box}$. Building on the judgmental reconstruction of
modal logic by \citet{Pfenning:2001aa}, \citet*{Nanevski:2008aa}
restrict the necessity modality to an explicit context and obtain
\emph{Contextual Modal Type Theory} (CMTT). A contextual code type
$[\Psi]A$ lists the context $\Psi$ in which its code is well typed, so
open code becomes expressible, and using the code in another context
requires an explicit substitution~\cite{Abadi:1991fr} that provides a
term for every variable in $\Psi$. Contextual modal types thus give a
uniform account of code with free variables and of the substitutions
that instantiate those variables.
CMTT has been extended in directions relevant to our paper:
Moebius~\cite{Jang:2022aa} combines it with System~F and pattern matching on
code. \LForallBox~\cite{murase2023contextual} adds polymorphic contexts.
Contextual MetaML~\cite{yin2026contextual} adds higher-order references that may
hold open code and establishes full abstraction via operational game semantics.
Closely related to our work, \citet{murase2023contextual} translate
$\lambda^{\bigcirc}$, restricted to two levels, into \LForallBox.
There are two insights from their translation that we use in our
compilation from EC to CMTT. First, quote and splice are translated by
constructing code contexts and explicit substitutions,
respectively. Second, code contexts are made explicit by following the
binding structure of generated code.

Two other type systems index code types by the typing context, independently of
CMTT. \citet*{Kim:2006cr} let the surrounding context capture variables by
name, so code may be spliced into any context that binds at least its free
variables, whereas the $\lambda^{[\,]}$ calculus of~\citet{rhiger2012staged}
binds them lexically.

Both EC and CMTT record information about the scope of open code in the code
type and support manipulating open code in the presence of mutable references.
However, they differ in how code variables are bound, as observed
by~\citet{murase2026refined}. We elaborate on the difference using examples.

\textbf{Variable binding in environment classifiers.}
Consider the following program in a metalanguage with environment
classifiers. It assembles two pieces of code, \CodeInline@c1@ and
\CodeInline@c2@, and splices them into a third. We write \CodeInline@``...''α@
for a quotation with classifier \ECalpha~and a tilde (\CodeInline@~@) for a
splice. Variable bindings are highlighted.
\Needspace{5\baselineskip}
\begin{lstlisting}[style=examplecode,lineskip=6pt]
  ``(λ (|\CodeListingVarMark{Salmon!60}{ec-bind-x}{x}|:Int) α. λ (|\CodeListingVarMark{SkyBlue!60}{ec-bind-y}{y}|:Int) β.     |\CodeListingBadge{Environment classifiers (EC)}|
      ~(let c1 = ``|\CodeListingVarMark{Salmon!60}{ec-occ-x1}{x}| * 2''α in  /* the scope of α contains x */
        let c2 = ``|\CodeListingVarMark{Salmon!60}{ec-occ-x2}{x}| + |\CodeListingVarMark{SkyBlue!60}{ec-occ-y}{y}|''β in  /* the scope of β contains x and y */
        ``~c1 - ~c2''β))
    43 1''ε  |\CodeListingArrowVia{RedOrange}{ec-occ-x2.north}{ec-bind-x.south}{right=24}{0.5}{ec-join-x}\CodeListingLine{RedOrange}{ec-occ-x1.west}{ec-join-x}{left=10}\CodeListingArrow{NavyBlue}{ec-occ-y.north}{ec-bind-y.south}{left=30}|
\end{lstlisting}
In a calculus with environment classifiers, a code variable is bound by a
$\lambda$-abstraction in its lexical scope, even though quotations and splices
intervene. The occurrences of \CodeInline@x@ in \CodeInline@c1@ and
\CodeInline@c2@ are bound by \CodeInline@λ (x:Int) α@ (red arrow), and the
\CodeInline@y@ in \CodeInline@c2@ by \CodeInline@λ (y:Int) β@ (blue arrow).
This binding structure is inherited from
$\lambda^{\bigcirc}$~\cite{davies1996temporal,Davies:2017aa}. Scopes of
binders in code are identified by names such as \ECalpha{} and \ECbeta. The
abstraction \CodeInline@λ (x:Int) α@ binds both the code variable \CodeInline@x@
and the classifier \ECalpha{} that names its scope (line 1), and the inner
abstraction \CodeInline@λ (y:Int) β@ binds \CodeInline@y@ and \ECbeta. The scope
named by \ECbeta{} lies within the scope named by \ECalpha, so the type system
introduces the subtyping $\ECalpha \sub \ECbeta$.
Both quotes and their code types are annotated with classifiers,
so \CodeInline@c1@ (line 2) has type \CodeInline@``Int''α@ and \CodeInline@c2@
has type \CodeInline@``Int''β@.
Due to the subtyping in this example, a piece of code with
classifier \ECalpha{} may be spliced wherever
classifier \ECalpha{} or \ECbeta{} is expected such as
the splice of \CodeInline@c1@ (line 4),
and \ECbeta{} may be spliced wherever
\ECbeta{} is expected, such as \CodeInline@c2@ (line 3).
A calculus based on EC thus relates
scopes through their names and the subtyping between those names.
We therefore refer to environment classifiers as a \emph{nominal} approach
to ensuring scope safety.
The type system ensures that metaevaluation ultimately produces closed code,
in this example
\CodeInline@``(λ (x:Int) α. λ (y:Int) β. (x * 2) - (x + y)) 43 1''ε@, where
\ECeps{} is the classifier of the empty scope.

\textbf{Variable binding in contextual modal types.}
Here is the same program in a metalanguage with contextual modal types. Variable
bindings are highlighted.
\Needspace{6\baselineskip}
\begin{lstlisting}[style=examplecode,lineskip=8pt]
  ``∅ ⊢ (λ (x:Int). λ (y:Int).  |\CodeListingBadge{Contextual modal types (CMTT)}|
         ~(let c1 = ``|\CodeListingVarMark{Salmon!60}{cm-bind-x1}{x}|:Int ⊢ |\CodeListingVarMark{Salmon!60}{cm-occ-x1}{x}| * 2'' in
           let c2 = ``|\CodeListingVarMark{Yellow!60}{cm-bind-x2}{x}|:Int, |\CodeListingVarMark{SkyBlue!60}{cm-bind-y}{y}|:Int ⊢ |\CodeListingVarMark{Yellow!60}{cm-occ-x2}{x}| + |\CodeListingVarMark{SkyBlue!60}{cm-occ-y}{y}|'' in
           ``x:Int, y:Int ⊢ ~c1{x=x} - ~c2{x=x,y=y}''){x=x,y=y})
        43 1''  |\CodeListingArrow{RedOrange}{cm-occ-x1.north}{cm-bind-x1.north}{right=35}\CodeListingArrow{BurntOrange}{cm-occ-x2.north}{cm-bind-x2.north}{right=20}\CodeListingArrow{NavyBlue}{cm-occ-y.south}{cm-bind-y.south}{left=25}|
\end{lstlisting}
In a calculus with contextual modal types, each piece of code carries a typing
context, such as \CodeInline@x:Int@ and \CodeInline@(x:Int, y:Int)@, which
identifies the scope of the code by spelling out the variables that the code
may mention together with their types. In CMTT, it is this context
annotation on a quotation that binds the variables of the quoted code, not an
enclosing $\lambda$-abstraction. The annotation \CodeInline@x:Int@ of
\CodeInline@c1@ binds the \CodeInline@x@ in \CodeInline@x * 2@ (red arrow, line
2), and the annotation \CodeInline@(x:Int, y:Int)@ of \CodeInline@c2@ binds
both \CodeInline@x@ (yellow arrow) and \CodeInline@y@ (blue arrow) in
\CodeInline@x + y@ (line 3). The \CodeInline@x@ in \CodeInline@c1@, the
\CodeInline@x@ in \CodeInline@c2@, and the \CodeInline@x@ bound by
\CodeInline@λ (x:Int)@ (line 1) are therefore three different variables, so
every splice must say how the variables of the spliced code relate to those of
the surrounding context. The splice \CodeInline@~c1{x=x}@ (line 4) applies to
\CodeInline@c1@ an explicit substitution of type
$\SubstTy{(x{:}\Int)}{(x{:}\Int,\, y{:}\Int)}$, which maps the \CodeInline@x@ of
\CodeInline@c1@, bound by its context annotation, to the \CodeInline@x@ of the
surrounding context \CodeInline@(x:Int, y:Int)@. Similarly, the splice
\CodeInline@~c2{x=x,y=y}@ substitutes the two variables of \CodeInline@c2@, and
the outermost splice applies \CodeInline@{x=x,y=y}@, which maps the variables
of the annotation on line 4 to the ones bound by the two
$\lambda$-abstractions. The program makes each substitution explicit, and the
type system checks that it provides a term for every variable of the context.
Whether code of one context can be used in another thus depends only on the
variable names and their types.
We therefore call CMTT the
\emph{structural} approach to scopes. Code types mention contexts, so
\CodeInline@c1@ has type \CodeInline@``x:Int ⊢ Int''@ and \CodeInline@c2@ has
type \CodeInline@``x:Int, y:Int ⊢ Int''@. Metaevaluation performs the
substitutions and produces the closed code
\CodeInline@``∅ ⊢ (λ (x:Int). λ (y:Int). (x * 2) - (x + y)) 43 1''@, where
\CodeInline@∅@ is the empty context.

In CMTT, code fragments can be assembled even if their scopes are not nested,
which EC cannot express. In the following program,
\CodeInline@c@ is constructed outside any $\lambda$-abstraction in code and
carries the context \CodeInline@y:Int@, which binds \CodeInline@y@ (line 1,
highlighted), and is then spliced under the abstraction \CodeInline@λ (x:Int)@.
\Needspace{4\baselineskip}
\begin{lstlisting}[style=examplecode,framextopmargin=3pt,framexbottommargin=6pt]
  let c = ``|\CodeListingVarMark{SkyBlue!60}{rc-bind-y}{y}|:Int ⊢ |\CodeListingVarMark{SkyBlue!60}{rc-occ-y}{y}| * 2'' in  |\CodeListingBadge{Reconnecting binders in CMTT}|
  ``∅ ⊢ (λ (|\CodeListingVarMark{Salmon!60}{rc-bind-x}{x}|:Int). ~c{y=|\CodeListingVarMark{Salmon!60}{rc-occ-x}{x}|}) 21''
\end{lstlisting}
\CodeListingArrow{NavyBlue}{rc-occ-y.north}{rc-bind-y.north}{right=20}%
\CodeListingArrow{RedOrange}{rc-occ-x.south}{rc-bind-x.south}{left=20}%
The substitution \CodeInline@{y=x}@ reconnects the variable \CodeInline@y@ of
\CodeInline@c@ to the variable \CodeInline@x@ bound by the abstraction (line 2,
highlighted), and metaevaluation produces the closed code
\CodeInline@``∅ ⊢ (λ (x:Int). x * 2) 21''@.

This program has no counterpart in EC, where the code
\CodeInline@y * 2@ can only be written within the scope of a $\lambda$ that
binds \CodeInline@y@ and can only be spliced into code within that scope. It can
never be constructed outside any $\lambda$ and rewired to a binder of an
unrelated scope. The difference between the two approaches mirrors the one
between nominal and structural type systems~\cite{Pierce:2002hj}: a nominal type
system relates two types only when the program declares their names related,
whereas a structural one relates them whenever their types match.

Environment classifiers identify the scope of code by a name, whereas
contextual modal types identify it by the typing context. We observe
that a classifier can be compiled into a code-level context that
includes the variables and types in scope at the classifier.
This intuition leads to the main contribution of our
paper: compiling EC into CMTT, and thereby connecting the nominal and
structural approaches to scopes.

\textbf{\textit{This paper:} compiling from EC to CMTT.} Our compilation extends
the translation of \citet{murase2023contextual}, discussed above, to a source
language with environment classifiers and mutable references. The source
language, \LAlpha, is a metaprogramming calculus with refined environment
classifiers and mutable references, extended with unrestricted subtyping and
subtype-bounded classifier polymorphism.
The target language, \LCtx, combines features of recent extensions of contextual
modal type theory. The \LCtx calculus is a two-level metaprogramming language
that supports mutable references and polymorphic contexts. To these, \LCtx adds
first-class substitutions in support of the compilation.
%
Both \LAlpha and \LCtx are restricted to two levels, because going beyond
two levels with mutable references in CMTT is an open
problem~\cite{yin2026contextual}. We do not include \texttt{run}, because
mutable references already make scope safety nontrivial, and the mechanism that
tracks open code fragments for typing \texttt{run} is analogous to the one for
references. We define a type-directed compilation from \LAlpha to \LCtx and
prove that it preserves types and that source and target programs generate the
same code. This \emph{connects the nominal scopes of EC to the structural
contexts of CMTT}, and it is the main contribution of this paper.

The key to our compilation is identifying the code-level context that an EC
corresponds to. The following program and its compilation illustrate this idea.
\par\noindent
\begin{minipage}{\linewidth}
\begin{lstlisting}[style=examplecode,numbers=none]
  ``(λ (x:Int) |\CodeListingVarBox{Salmon!60}{\CodeListingEC{Plum}{\textalpha}}|. λ (y:Int) |\CodeListingVarMark{Yellow!60}{cx-bind-b}{\CodeListingEC{MidnightBlue}{\textbeta}}|.  |\CodeListingBadge{Source: EC}|
       ~(let c = ``|\CodeListingFrameBox{RedOrange}{x}| + y''|\CodeListingVarMark{Yellow!60}{cx-occ-b}{\CodeListingEC{MidnightBlue}{\textbeta}}|
         in c))
    41 1''|\CodeListingVarBox{SkyBlue!60}{\CodeListingEC{Turquoise}{\textepsilon}}\CodeListingArrow{BurntOrange}{cx-occ-b.south}{cx-bind-b.south}{right=40}|
\end{lstlisting}
\vspace{-0.3\baselineskip}
\centerline{\textit{\small compiles to} $\boldsymbol{\compile}$}
\begin{lstlisting}[style=examplecode,numbers=none]
  ``|\CodeListingVarBox{SkyBlue!60}{\CodeListingEmptyCtx}| ⊢   |\CodeListingBadge{Target: CMTT}|
    (λ (x:Int). λ (y:Int).
       ~(let c = ``|\CodeListingVarBox{Yellow!60}{x:Int, y:Int}| ⊢ |\CodeListingFrameBox{RedOrange}{\textasciitilde\CodeQuoteOpen\CodeListingVarBox{Salmon!60}{x:Int} \CodeListingTurnstile{} x\CodeQuoteClose\{x=x\}}| + y''
         in c){x=x,y=y})
    41 1''
\end{lstlisting}
\end{minipage}
\par\noindent
For each code-level binder annotated with a classifier, such as \ECbeta,
the compilation recovers the context $\Gamma_{\ECbeta}$ of the scope that the
binder introduces by following EC subtyping outwards. The binder lies in the
scope of an enclosing classifier \ECalpha, so $\Gamma_{\ECbeta}$ is $\Gamma_{\ECalpha}$
extended with the variable that the binder binds, and the chain ends at the
empty classifier \ECeps, whose context is empty. The compiler calculates the context
for each classifier: $\Gamma_{\ECalpha} = x{:}\Int$,
$\Gamma_{\ECbeta} = (x{:}\Int, y{:}\Int)$, and $\Gamma_{\ECeps} = \emptyset$.
Every classifier annotation of the source is then replaced by the
context that it corresponds to, shown in matching colors:
\CodeHighlight{Salmon!60}{\ECalpha} corresponds to \CodeHighlight{Salmon!60}{x:Int}
and \CodeHighlight{Yellow!60}{\ECbeta} to \CodeHighlight{Yellow!60}{x:Int,
  y:Int}.
The quotation \CodeInline@``x + y''@\CodeHighlight{Yellow!60}{\ECbeta}
becomes a quotation annotated with \CodeHighlight{Yellow!60}{x:Int, y:Int}, and
the outer quotation, of the empty classifier \CodeHighlight{SkyBlue!60}{\ECeps},
is annotated with the empty context
\CodeHighlight{SkyBlue!60}{\CodeListingEmptyCtx}. The code-level abstractions
lose their classifiers, since the contexts now carry that information. The
interesting part is the framed occurrence of \CodeInline@x@, which is bound in
the scope named \ECalpha{} but used in code of classifier \ECbeta. This is
allowed because the scope named by \ECbeta{} lies within the scope named by
\ECalpha. In the target, this occurrence compiles to a quotation of
\CodeInline@x@ in its own context \CodeInline@x:Int@, spliced into the larger
context
\CodeInline@(x:Int, y:Int)@ under the substitution \CodeInline@{x=x}@. Scope
nesting therefore becomes context weakening, witnessed by a substitution.
Finally, the splice around the let-expression gains the substitution
\CodeInline@{x=x,y=y}@, which maps the two variables of the quotation's context
to the ones bound by the $\lambda$-abstractions.

\textbf{Metatheoretical results.} We prove in Agda that the compilation
preserves types and is semantically correct by a bisimulation between
configurations of the source and the target. Contextual modal types are thus at
least as expressive as environment classifiers, in that any scope described by a
name can also be described by a context. We conjecture that the converse
direction is difficult, because a classifier can only name a scope that arises
from nested binders in code, whereas a contextual modal type may mention any
context.

\textbf{Technical Contributions.}
This paper makes the following contributions.
\begin{enumerate}
\item We present \LCtx, a two-level CMTT calculus that adds
  first-class substitutions (to support the compilation) and combines
  support for mutable references~\cite{yin2026contextual} and
  polymorphic contexts~\cite{murase2023contextual}.
  
\item We define a type-directed compilation from \LAlpha to \LCtx, thereby
  establishing the connection between the nominal scopes of environment
  classifiers and the structural contexts of contextual modal types.
\item We prove that the compilation preserves types and that it is
  semantically correct: source and target programs generate the same code and
  agree on divergence. Both proofs are fully mechanized in Agda.
\end{enumerate}

Section~\ref{sec:examples} introduces the compilation through examples,
Sections~\ref{sec:source-language} and~\ref{sec:target-language} present
\LAlpha and \LCtx, Section~\ref{sec:compilation} defines the compilation,
Section~\ref{sec:metatheory} proves it type preserving and semantically correct,
and Section~\ref{sec:conclusion} concludes. The Appendix, the Agda proof
scripts, and the Haskell code for a prototype implementation of the compilation
are all in the supplementary material of this paper.

\section{From EC to CMTT, by Examples}
\label{sec:examples}

This section demonstrates the compilation from \LAlpha to \LCtx through
examples. We start with the basic translation of classifiers into contexts
(\S~\ref{sec:examples-basic}). We then compile the two forms of subtyping in
\LAlpha, between classifiers and between types
(\S~\ref{sec:examples-subtyping}), into substitutions and coercion functions in \LCtx.
Finally, we compile classifier
polymorphism and subtype constraints into context polymorphism and first-class
substitutions (\S~\ref{sec:examples-polymorphism}).

\subsection{Compiling Environment Classifiers to Contexts}
\label{sec:examples-basic}

In this section, we illustrate the basic pattern of our compilation, which
has two parts.
First, classifiers become contexts, so code types and quotations mention
contexts instead of classifiers. Second, splices become splices under
substitutions. We also preview the semantic correctness of the compilation.

Consider the following metaprogram, which generates the code of $x^4$ by
squaring twice.
\Needspace{5\baselineskip}
\begin{lstlisting}[style=examplecode]
  ``λ (x:Int) α.  |\CodeListingBadge{Source program (EC)}|
    ~(let square = λ (c:``Int''α). ``~c * ~c''α
      in square (square ``x''α))''ε
\end{lstlisting}
{\emergencystretch=3em The program uses a single classifier. The 
  $\lambda$-abstraction on line 1 binds \CodeInline@x@ and names its scope
  \ECalpha. The metafunction
\CodeInline@square@ (line 2) takes code of classifier \ECalpha{} and returns
code of the same classifier that multiplies the argument of the function by
itself. Applying \CodeInline@square@ twice to \CodeInline@``x''α@ (line 3) produces the code
\CodeInline@``λ (x:Int) α. (x * x) * (x * x)''ε@.\par}

The compilation first identifies the context of each classifier. The abstraction
on line 1 binds \CodeInline@x@ in the empty scope, so $\Gamma_{\ECalpha} = x{:}\Int$
and $\Gamma_{\ECeps} = \emptyset$. The compilation then replaces every
classifier by its context and equips every splice with a substitution. In the
listing below, contexts are highlighted in the color of their classifier, and
substitutions are underlined.
\Needspace{5\baselineskip}
\begin{lstlisting}[style=examplecode]
  ``|\CodeListingVarBox{CtxEps}{\CodeListingEmptyCtx}| ⊢ λ (x:Int).  |\CodeListingBadge{Target program (CMTT)}|
    ~(let square = λ (c:``|\CodeListingVarBox{CtxAlpha}{x:Int}| ⊢ Int''). ``|\CodeListingVarBox{CtxAlpha}{x:Int}| ⊢ ~c|\CodeListingUnderline{RedOrange}{\{x=x\}}| * ~c|\CodeListingUnderline{RedOrange}{\{x=x\}}|''
      in square (square ``|\CodeListingVarBox{CtxAlpha}{x:Int}| ⊢ x''))|\CodeListingUnderline{RedOrange}{\{x=x\}}|''
\end{lstlisting}
Code types and quotations now mention contexts instead of classifiers. The type
\CodeInline@``Int''α@ of the parameter \CodeInline@c@ becomes
\CodeInline@``@\CodeHighlight{CtxAlpha}{x:Int}\CodeInline@ ⊢ Int''@, and each
quotation \CodeInline@``...''α@ becomes
\CodeInline@``@\CodeHighlight{CtxAlpha}{x:Int}\CodeInline@ ⊢ ...''@. Every
splice applies a substitution (underlined). Here every quotation is spliced into
code of the same classifier \ECalpha, so every substitution is the identity on
$\Gamma_{\ECalpha}$, written \CodeInline@{x=x}@. The substitutions connect different
binders. On line 2, they map the \CodeInline@x@ of the spliced code, bound by
its context annotation \CodeInline{x:Int}, to the \CodeInline@x@ bound by the
context annotation of the surrounding quotation; on line 3, to the
\CodeInline@x@ bound by the $\lambda$-abstraction of line 1.

{\emergencystretch=3em
The target program metaevaluates to
\CodeInline@``∅ ⊢ λ (x:Int). (x * x) * (x * x)''@. This example shows
the basic pattern of the compilation: classifiers become contexts, and
splices become splices with substitutions. Both the source metaprogram
and the target produce \CodeInline@λ (x:Int). (x * x) * (x * x)@.  In
general, we prove that the target always generates the same code as
the source does with the semantic correctness theorem in
\S~\ref{sec:semantic-correctness}. \par}

\subsection{Compiling Subtyping}
\label{sec:examples-subtyping}

Code of an outer scope should be usable wherever code of an inner scope is
expected, and \LAlpha allows this through subtyping. From \LAlpha to \LCtx, we
compile subtyping away through a coercion
semantics~\cite{Breazu-Tannen:1991aa,Pierce:2002hj}.
In the following subsections, we show that subtyping between classifiers
compiles to a substitution, and subtyping between types compiles to a coercion
function that applies the substitution through a splice.


\textbf{Subtyping between classifiers.} In the example of
Section~\ref{sec:introduction}, a variable of an outer scope occurs in code of
an inner scope, and the compilation moves it there by splicing it under a
substitution and re-quoting it in the inner context. In general, a derivation of
$\ECalpha \sub \ECbeta$ compiles to a substitution of type
$\SubstTy{\Gamma_{\ECalpha}}{\Gamma_{\ECbeta}}$, where $\Gamma_{\ECalpha}$ and
$\Gamma_{\ECbeta}$ are the contexts that \ECalpha{} and \ECbeta{} compile to.
The following example demonstrates the compilation of subtyping between
classifiers.
\Needspace{5\baselineskip}
\begin{lstlisting}[style=examplecode]
  ``(λ (x:Int) α. λ (y:Int) β. λ (z:Int) γ.  |\CodeListingBadge{Source program (EC): subtyping}|
      ~``x + y''γ)
    1 41 0''ε
\end{lstlisting}
We have $\ECalpha \sub \ECbeta \sub \ECgamma$ from the three nested
$\lambda$-abstractions. The quotation \CodeInline@``x + y''γ@ (line 2) mentions
two variables of outer scopes. The variable \CodeInline@y@ is bound in the scope
named by \ECbeta, so its occurrence requires $\ECbeta \sub \ECgamma$. The
variable \CodeInline@x@ is bound in the scope named by \ECalpha, so its
occurrence requires $\ECalpha \sub \ECgamma$, which holds by transitivity
through \ECbeta. Here is the compilation of the program.
\Needspace{5\baselineskip}
\begin{lstlisting}[style=examplecode]
  ``∅ ⊢ (λ (x:Int). λ (y:Int). λ (z:Int).  |\CodeListingBadge{Target program (CMTT)}|
      ~``|\CodeListingVarBox{CtxGamma}{x:Int, y:Int, z:Int}| ⊢
           ~``|\CodeListingVarBox{CtxAlpha}{x:Int}| ⊢ x'' (|\CodeListingUnderline{RedOrange}{\{x=x\} \CodeListingCompose{} \{x=x,y=y\}}|) +
           ~``|\CodeListingVarBox{CtxBeta}{x:Int, y:Int}| ⊢ y'' {x=x,y=y}''
       {x=x,y=y,z=z})
    1 41 0''
\end{lstlisting}
The contexts are $\Gamma_{\ECalpha} = x{:}\Int$, $\Gamma_{\ECbeta} = (x{:}\Int,\,
y{:}\Int)$, and $\Gamma_{\ECgamma} = (x{:}\Int,\, y{:}\Int,\, z{:}\Int)$. The
occurrence of \CodeInline@y@ is spliced under the substitution
\CodeInline@{x=x,y=y}@, which is of type $\SubstTy{\Gamma_{\ECbeta}}{\Gamma_{\ECgamma}}$
(line 4, underlined). Transitivity of subtyping compiles to the composition of
substitutions. The occurrence of \CodeInline@x@ is spliced under the
substitution
\CodeInline@{x=x} ∘ {x=x,y=y}@ (line 3, underlined), which composes the
substitution for $\ECalpha \sub \ECbeta$ with the one for $\ECbeta \sub
\ECgamma$ and thus has type $\SubstTy{\Gamma_{\ECalpha}}{\Gamma_{\ECgamma}}$.

\textbf{Subtyping between types.} The following example demonstrates subtyping
between types by storing code of classifier \ECalpha{} into a reference cell
that expects code of classifier \ECbeta. We have $\ECalpha \sub \ECbeta$ from the
nesting of $\lambda$-abstractions.
\Needspace{6\baselineskip}
\begin{lstlisting}[style=examplecode]
  ``(λ (x:Int) α. λ (y:Int) β.  |\CodeListingBadge{Source program (EC): subtyping between types}|
      ~(let r = ref ``0''β in  /* r : Ref ``Int''β */
        r := ``x''α;           /* ``Int''α <: ``Int''β */
        !r))
    42 24''ε
\end{lstlisting}
The reference \CodeInline@r@ holds code of classifier \ECbeta{} and has type
\CodeInline@Ref ``Int''β@ (line 2). Line 3 assigns to it the code
\CodeInline@``x''α@, which the type system accepts by subsumption.
Line 4 reads the code back and splices it into the inner abstraction.
Here is the compilation of the program.
\Needspace{7\baselineskip}
\begin{lstlisting}[style=examplecode]
  ``∅ ⊢ (λ (x:Int). λ (y:Int).  |\CodeListingBadge{Target program (CMTT)}|
      ~(let r = ref ``|\CodeListingVarBox{CtxBeta}{x:Int, y:Int}| ⊢ 0'' in
        r := |\CodeListingFrameBox{RedOrange}{(\CodeListingLambda{} (c:\CodeQuote{\CodeListingVarBoxRaw{CtxAlpha}{x:Int} \CodeListingTurnstile{} Int}). \CodeQuote{\CodeListingVarBoxRaw{CtxBeta}{x:Int, y:Int} \CodeListingTurnstile{} \textasciitilde{}c\{x=x\}})}|
               ``|\CodeListingVarBox{CtxAlpha}{x:Int}| ⊢ x'';
        !r){x=x,y=y})
    42 24''
\end{lstlisting}
The reference holds code of type
\CodeInline@``@\CodeHighlight{CtxBeta}{x:Int, y:Int}\CodeInline@ ⊢ Int''@ (line
2), whereas the code assigned to it, \CodeInline@``x:Int ⊢ x''@ (line 4), has
type \CodeInline@``@\CodeHighlight{CtxAlpha}{x:Int}\CodeInline@ ⊢ Int''@. The
highlighted contexts correspond to \ECbeta{} and \ECalpha. To convert between
the two types, the compilation inserts the framed function on line 3, which is
the coercion that \CodeInline@``Int''α <: ``Int''β@ compiles to. It splices its
argument under the substitution \CodeInline@{x=x}@ and re-quotes the result in
the context
\CodeInline@(x:Int, y:Int)@. In general, the coercion between two code types
applies the substitution between their classifiers through a splice and
re-quotes the result, and the compilation inserts it wherever the typing
derivation of the source uses subsumption.



\subsection{Compiling Bounded Quantification for Classifiers}
\label{sec:examples-polymorphism}

\LAlpha expresses bounded quantification for classifiers with two features,
classifier polymorphism and subtype constraints. We compile these language
features separately. Classifier abstraction and application compile to context
abstraction and application. Subtype constraint introduction and elimination
compile to abstraction over and application to a first-class substitution,
respectively.

A metafunction that manipulates code of a fixed classifier, such as
\CodeInline@square@ in \S~\ref{sec:examples-basic}, can only be used in one
scope. For reuse across scopes, \LAlpha supports classifier polymorphism, an
idea that \citet{Kiselyov:2016aa} sketch informally and
\citet{murase2026refined} formalize.
\CodeInline@(Λ α. M)@ abstracts over a classifier, and \CodeInline@M[β]@
instantiates the abstraction with \ECbeta. Subtype constraints let a
polymorphic metafunction mention the variables of a specific scope. For example, in the term
\CodeInline@β <: γ => M@, where \ECgamma{} is a polymorphic classifier and
\ECbeta{} the name of a scope, the constraint $\ECbeta \sub \ECgamma$ lets code
of classifier \ECgamma{} in \CodeInline@M@ use the variables of the scope named
by \ECbeta. The elimination form \CodeInline@M •@ discharges the subtype
constraint in the type of
\CodeInline@M@ by checking that the subtyping holds.

The following example uses bounded quantification for classifiers, expressed as
a classifier abstraction followed by a subtype constraint.
\Needspace{6\baselineskip}
\begin{lstlisting}[style=examplecode]
  ``(λ (x:Int) α. λ (y:Int) β.  |\CodeListingBadge{Source program (EC): bounded quantification}|
      ~(let l = |\CodeListingFrameBox{Goldenrod}{\CodeListingBigLambda{} \CodeListingEC{PineGreen}{\textgamma}.}| |\CodeListingFrameBox{RedOrange}{\CodeListingEC{MidnightBlue}{\textbeta} <: \CodeListingEC{PineGreen}{\textgamma} \CodeListingDoubleFun{}}| ``x + y''γ in
        l |\CodeListingFrameBox{Green}{[\CodeListingEC{MidnightBlue}{\textbeta}]}| |\CodeListingFrameBox{NavyBlue}{\CodeListingBullet{}}| ))
    42 0''ε
\end{lstlisting}
The function \CodeInline@l@ (line 2) generates the code \CodeInline@x + y@ at
the polymorphic classifier \ECgamma. Its body mentions \CodeInline@x@ and
\CodeInline@y@, bound in the scopes named by \ECalpha{} and \ECbeta, and both
are visible in code of classifier \ECgamma{} thanks to the constraint $\ECbeta
\sub \ECgamma$. Line 3 instantiates \CodeInline@l@ at \ECbeta{} and discharges
the constraint, which reads $\ECbeta \sub \ECbeta$ after instantiation and
holds by reflexivity. The result has type \CodeInline@``Int''β@ and is spliced
into the inner abstraction of the same classification.

Here is the compilation. The frames of matching colors relate the constructs of
the source to those of the target. We explain the compilation below.
\Needspace{7\baselineskip}
\begin{lstlisting}[style=examplecode]
  ``∅ ⊢ (λ (x:Int). λ (y:Int).  |\CodeListingBadge{Target program (CMTT): context polymorphism}|
      ~(let l = |\CodeListingFrameBox{Goldenrod}{\CodeListingBigLambda{} \CodeListingEC{PineGreen}{\textgamma}.}| |\CodeListingFrameBox{RedOrange}{\CodeListingLambda{} (s:(x:Int, y:Int) \CodeListingSubstFun{} \CodeListingVarBoxRaw{CtxGamma}{\CodeListingEC{PineGreen}{\textgamma}}).}|
                 ``|\CodeListingVarBox{CtxGamma}{\CodeListingEC{PineGreen}{\textgamma}}| ⊢ ~``x:Int ⊢ x'' ({x=x} ∘ s) +
                        ~``x:Int, y:Int ⊢ y'' s'' in
        l |\CodeListingFrameBox{Green}{[x:Int, y:Int]}| |\CodeListingFrameBox{NavyBlue}{\{x=x,y=y\}}|){x=x,y=y})
    42 0''
\end{lstlisting}

\textbf{Compiling classifier polymorphism.} Classifier
abstraction compiles to context abstraction (line 2, yellow box), and classifier
application to context application (line 5, green box). To support the
compilation, \LCtx supports context polymorphism. The polymorphic classifier
\ECgamma{} compiles to a context variable of the same name, so the quotation
\CodeInline@``x + y''γ@ is now in the context \CodeInline@γ@ (line 3,
highlighted). The instantiation at \ECbeta{} becomes an instantiation at the
context of \ECbeta, which is \CodeInline@(x:Int, y:Int)@.

\textbf{Compiling subtype constraints.} Recall from
\S~\ref{sec:examples-subtyping} that the coercion for subtyping between
classifiers is a substitution. Consequently, a subtype constraint introduction
compiles to a function that takes a substitution (line 2, red box). The function
takes the substitution for the subtyping $\ECbeta \sub \ECgamma$, so the
argument is of type \CodeInline@(x:Int, y:Int) ⇛ γ@. A constraint elimination
compiles to a function application. In the example, the application passes the
substitution for the subtyping that the elimination checks, which is $\ECbeta
\sub \ECbeta$ after instantiation, so the argument is the identity substitution
\CodeInline@{x=x,y=y}@ (line 5, blue box).


In summary, bounded quantification for classifiers compiles to context
polymorphism and to abstraction over and application to first-class
substitutions. To support compilation, substitutions must be first-class
citizens in \LCtx.

\section{Source Language: \LAlpha{} with Environment Classifiers}
\label{sec:source-language}

In this section, we define \LAlpha, the source language of the compilation.
\LAlpha is a two-level calculus with refined environment
classifiers~\cite{Kiselyov:2016aa} and mutable references, so a code type
records the classifier that names the scope in which the code is meaningful, and
a reference may hold open code. To these, \LAlpha adds unrestricted subtyping
between types, and classifier polymorphism with bounded quantification in the
style of R$^2$EC~\cite{murase2026refined}. We define the syntax in
\S~\ref{sec:lambda-alpha-ref-syntax}, the type system in
\S~\ref{sec:lambda-alpha-ref-typesystem}, and the operational semantics in
\S~\ref{sec:lambda-alpha-ref-semantics}.

\subsection{Syntax of \LAlpha{}}\label{sec:lambda-alpha-ref-syntax}

\begin{figure}[tbp]
\footnotesize
\setlength{\abovedisplayskip}{2pt}\setlength{\belowdisplayskip}{2pt}%
\[
\begin{array}{lrcl}
\text{base types}        & \iota & ::= & \Int \mid \Unit \\
\text{code types}        & S, T & ::= & \iota \mid S \to T \\
\text{EC variables}      & \ECalpha,\ECbeta,\ECgamma \\
\text{ECs}               & e  & ::= & \ECalpha \mid \ECeps \\
\text{meta types}        & A,B & ::= & \iota \mid A \to B \mid \REFTY{A} \mid \CodeQuote{T}e
                           \mid \forall \ECalpha.\,A \mid e{<:}e{\Rightarrow}A \\
\text{contexts}          & \Gamma & ::= & \emptyset \mid \Gamma,\ECalpha
                           \mid \Gamma,e_1{<:}e_2
                           \mid \Gamma,x:A \mid \Gamma,(x:T)^{\ECalpha}
\\
\text{subtype contexts} & \Theta & ::= & \emptyset \mid \Theta,e_1{<:}e_2
\\
\text{constants}  & k \\
\text{term variables}  & x,y,z,... \\
\text{code terms} & L^\circ,M^\circ,N^\circ &::=&
  k \mid x \mid \lambda (x{:}T)^{\ECalpha}.\,M^\circ
  \mid M^\circ\;N^\circ \mid \SPLICE{M}\\
\text{meta terms} & L,M,N &::=&
  k \mid x \mid \lambda x{:}A.\,M \mid M\;N \\
&&& \mid \Lambda \ECalpha.\,M \mid M[e]
  \mid e_1{<:}e_2{\Rightarrow}M \mid M \bullet \\
&&& \mid \CodeQuote{M^\circ}e \mid \REF{M} \mid M := N \mid \mathop{!}M
\end{array}
\]
\caption{Syntax of \LAlpha{}.}
\label{fig:lambda-alpha-ref-syntax}
\end{figure}

Figure~\ref{fig:lambda-alpha-ref-syntax} defines the syntax of \LAlpha. It
generates code in STLC, and the metalanguage uses classifiers to ensure that the
generated code is well scoped in the presence of mutable references. We describe
the EC-specific features below.

\textbf{Code and environment classifiers.} Following \citet{Taha:2003qo}, a
quotation $\CodeQuote{M^\circ}e$ and a code type $\CodeQuote{T}e$ carry an
environment classifier $e$, which represents the scope in which the code is
meaningful. A classifier is either \ECeps, which is the name of the empty scope,
or a classifier variable \ECalpha. Following the refined environment classifiers
of \citet{Kiselyov:2016aa}, a code-level abstraction $\lambda
(x{:}T)^{\ECalpha}.\,M^\circ$ binds both its parameter \CodeInline@x@ and the
classifier \ECalpha{} that gives the scope of its body a name. A splice
$\SPLICE{M}$ integrates the code that the meta term $M$ produces into the
surrounding code.

\textbf{Bounded quantification for classifiers.} Following
R$^2$EC~\cite{murase2026refined}, \LAlpha supports bounded quantification over
classifiers, which it expresses with two features, classifier polymorphism and
subtype constraints. The universal type $\forall \ECalpha.\,A$ quantifies over
classifiers. Classifier abstraction $\Lambda \ECalpha.\,M$ binds a classifier
variable (\ECalpha), whereas classifier application $M[e]$ instantiates it with
a classifier ($e$). A subtype-constrained type $e_1{<:}e_2{\Rightarrow}A$
introduces the constraint $e_1 <: e_2$ in the body type $A$. The subtype
constraint introduction form $e_1{<:}e_2{\Rightarrow}M$ assumes the constraint
while typing $M$, and its elimination form $M\bullet$ discharges the constraint
where the subtyping holds.

\subsection{Type System of \LAlpha{}}
\label{sec:lambda-alpha-ref-typesystem}

\begin{figure}[tbp]
\footnotesize
\hfill\Fbox{$\Theta \vdash e_1 <: e_2$}
\begin{gather*}
  \inference{e_1 <: e_2 \in \Theta}
            {\Theta \vdash e_1 <: e_2}
  \quad
  \inference{}
            {\Theta \vdash \ECeps <: e}
  \quad
  \inference{}
            {\Theta \vdash e <: e}
  \quad
  \inference{\Theta \vdash e_1 <: e_2
             & \Theta \vdash e_2 <: e_3}
            {\Theta \vdash e_1 <: e_3}
\end{gather*}

\hfill\Fbox{$\Gamma;e \vdash M^\circ : T$}
\vspace{-10pt}
\begin{gather*}
  \inference{(x:T)^{\ECalpha} \in \Gamma}
            {\Gamma;\ECalpha \vdash x : T}
  \RuleTag{${\vdash}$code-var}
  \quad
  \inference{\Gamma \vdash M : \CodeQuote{T}e}
            {\Gamma;e \vdash \SPLICE{M} : T}
  \RuleTag{${\vdash}$splice}
  \\[0.7ex]
  \inference{\Gamma,\RuleHighlight{\ECalpha,e{<:}\ECalpha,(x:S)^{\ECalpha}};\ECalpha
             \vdash N^\circ : T
            \\ \ECalpha \notin \Gamma}
            {\Gamma;e \vdash \lambda (x{:}S)^{\ECalpha}.\,N^\circ : S \to T}
  \RuleTag{${\vdash}$code-abs}
  \quad
  \inference{\Gamma;e_1 \vdash M^\circ : T
            \\ \RuleHighlight{\mathsf{Sub}(\Gamma) \vdash e_1 <: e_2}}
            {\Gamma;e_2 \vdash M^\circ : T}
  \RuleTag{${\vdash}$code-subenv}
  \\ \cdots
\end{gather*}

\hfill\Fbox{$\Gamma \vdash M : A$}
\vspace{-20pt}
\begin{gather*}
  \inference{\Gamma \vdash e
             \\
             \Gamma;e \vdash M^\circ : T}
            {\Gamma \vdash \CodeQuote{M^\circ}e : \CodeQuote{T}e}
  \RuleTag{${\vdash}$quote}
  \quad
  \inference{\Gamma \vdash e_1
             &
             \Gamma \vdash e_2
             \\
             \Gamma,\RuleHighlight{e_1{<:}e_2} \vdash M : A}
            {\Gamma \vdash e_1{<:}e_2{\Rightarrow}M :
             e_1{<:}e_2{\Rightarrow}A}
  \RuleTag{${\vdash}$sub-intro}
  \\[0.7ex]
  \inference{\Gamma \vdash M : e_1{<:}e_2{\Rightarrow}A &
             \RuleHighlight{\mathsf{Sub}(\Gamma) \vdash e_1 <: e_2}}
            {\Gamma \vdash M\bullet : A}
  \RuleTag{${\vdash}$sub-elim}
  \qquad \cdots
\end{gather*}
\caption{Selected typing rules of \LAlpha{} (excerpt of
  Figure~\ref{fig:lambda-alpha-ref-typing-full} of the Appendix).}
\label{fig:lambda-alpha-ref-type-system}
\end{figure}

Figure~\ref{fig:lambda-alpha-ref-type-system} shows selected typing rules
of \LAlpha; the complete rules are in
Figure~\ref{fig:lambda-alpha-ref-typing-full} of the Appendix. All judgments
carry a typing context $\Gamma$, which binds classifiers, code and meta
variables, and records subtype constraints. The typing judgment for meta terms
takes the form $\Gamma \vdash M : A$. The typing judgment for code terms,
$\Gamma;e \vdash M^\circ : T$, also carries the classifier $e$ of the nearest
enclosing scope. The two judgments are mutually recursive through quotation and
splice. Well-formedness and subtyping of types are straightforward.
They are defined in Figures~\ref{fig:lambda-alpha-ref-well-formedness}
and~\ref{fig:lambda-alpha-ref-subtyping-full} of the Appendix, respectively.

\textbf{Subtyping between classifiers.} The subtyping judgment between
classifiers, $\Theta \vdash e_1 <: e_2$, is defined in
Figure~\ref{fig:lambda-alpha-ref-type-system}. It derives subtyping from a
subtype context $\Theta$, which $\mathsf{Sub}(\Gamma)$ extracts from the subtype
constraints in $\Gamma$. The constraints in $\Theta$ form the edges of a graph,
with an edge from \ECeps{} to every classifier. The subtyping $e_1 <: e_2$ holds
when the graph has a path from $e_1$ to $e_2$.

\textbf{Typing code terms.} The typing rule for code variables
(${\vdash}$\textit{code-var}) enforces that a variable may only occur where the
nearest enclosing scope has the same classifier. The rule for code
$\lambda$-abstractions (${\vdash}$\textit{code-abs}) binds both a code variable
\CodeInline@x@ and the classifier \ECalpha{}. It also links \ECalpha{} to the
enclosing scope by the constraint $e <: \ECalpha$ and types the body of the
$\lambda$ under the extended context (highlighted). A splice
(${\vdash}$\textit{splice}) types its meta term, which must produce code of the
enclosing classifier. The subsumption rule between classifiers
(${\vdash}$\textit{code-subenv}) lets code typed under $e_1$ be used in the
scope represented by $e_2$ when $e_1 <: e_2$ holds (highlighted).

\textbf{Typing meta terms.} A quotation (${\vdash}$\textit{quote}) types its
code term with the annotated classifier $e$ as the enclosing scope. Classifier
abstraction and application follow System~F with classifiers in place of types,
and their rules are in the Appendix. Subtype constraint introduction
(${\vdash}$\textit{sub-intro}) extends $\Gamma$ with the assumption $e_1 <: e_2$
(highlighted), and subtype constraint elimination (${\vdash}$\textit{sub-elim})
discharges the assumption by checking that $e_1 <: e_2$ holds (highlighted).

\subsection{Operational Semantics of \LAlpha{}}\label{sec:lambda-alpha-ref-semantics}

\begin{figure}[tbp]
\footnotesize
\setlength{\abovedisplayskip}{2pt}\setlength{\belowdisplayskip}{2pt}%
\[
\begin{array}{lrcl}
  \text{addresses}         & a & \in & \mathbb{N} \\
  \text{heap typing contexts} & \Sigma & ::= & \emptyset \mid \Sigma, a{:}A \\
  \text{runtime code terms} & L^\circ,M^\circ,N^\circ &::=&
    \dots \mid \overline{\lambda} (x{:}T)^{\ECalpha}.\,M^\circ \\
  \text{runtime meta terms} & L,M,N &::=& \dots \mid a \\
  \text{code values} & V^\circ & ::= &
     k \mid x \mid \overline{\lambda} (x{:}T)^{\ECalpha}.\,V^\circ
     \mid V^\circ\;V^\circ\\
  \text{meta values} & V & ::= &
     k \mid \lambda x{:}A.\,M \mid a \mid \CodeQuote{V^\circ}e
     \mid \Lambda \ECalpha.\,M \mid e{<:}e{\Rightarrow}M\\[0.5ex]
  \text{heaps} & \mu & ::= & \emptyset \mid \mu[a\mapsto V]
\end{array}
\]

\hfill\Fbox{$\Gamma;\Sigma;e \vdash M^\circ : T$}
\vspace{-5pt}
\begin{gather*}
  \inference{\RuleHighlight{\Gamma \vdash \ECalpha}
             & \RuleHighlight{(x:S)^{\ECalpha} \in \Gamma}
             & \RuleHighlight{e <: \ECalpha \in \Gamma}
             & \Gamma;\Sigma;\ECalpha \vdash N^\circ : T}
            {\Gamma;\Sigma;e \vdash \overline{\lambda} (x{:}S)^{\ECalpha}.\,N^\circ : S \to T}
            \RuleTag{${\vdash}$code-abs-val}
            \\\cdots
\end{gather*}
\hfill\Fbox{$\Gamma;\Sigma \vdash M : A$}
\vspace{-10pt}
\begin{gather*}
  \inference{a{:}A \in \Sigma}
            {\Gamma;\Sigma \vdash a : \REFTY{A}}
            \RuleTag{${\vdash}$addr}
            \qquad\cdots
\end{gather*}

\hfill\Fbox{$
  e \vdash \langle\Gamma;M^\circ;\mu\rangle
  \longrightarrow^\circ
  \langle\Gamma';N^\circ;\mu'\rangle
  $}
\begin{gather*}
  \inference{\langle\Gamma;M;\mu\rangle
              \longrightarrow^m
              \langle\Gamma';N;\mu'\rangle}
            {e \vdash \langle\Gamma;\SPLICE{M};\mu\rangle
              \longrightarrow^\circ
              \langle\Gamma';\SPLICE{N};\mu'\rangle}
  \RuleTag{$\xi$-splice}
  \\[0.5ex]
  \inference{}
  {e \vdash \langle\Gamma;\SPLICE{\CodeQuote{V^\circ}e'};\mu\rangle
    \longrightarrow^\circ
    \langle\Gamma;V^\circ;\mu\rangle}
  \RuleTag{$\beta$-splice}
  \\[0.5ex]
  \inference{\ECbeta,y\notin\Gamma}
  {\begin{array}{c}
             e\vdash\langle \Gamma;\lambda(x{:}T)^{\ECalpha}.\,N^\circ;\mu\rangle
             \longrightarrow^\circ \\
             \langle (\Gamma,\RuleHighlight{\ECbeta,(e{<:}\ECbeta),
               (y{:}T)^{\ECbeta}});
               \overline{\lambda}(y{:}T)^{\ECbeta}.\,
               N^\circ[\ECalpha:=\ECbeta][x:=y];\mu\rangle
             \end{array}}
  \RuleTag{code-$\lambda$}
  \\[0.5ex]
  \inference{\ECalpha\vdash\langle \Gamma; 
               N^\circ;\mu\rangle 
               \longrightarrow^\circ 
               \langle \Gamma';N^{\circ\prime};\mu'\rangle
             }
            {
             e\vdash\langle \Gamma;
             \overline{\lambda}
             (x{:}T)^{\ECalpha}.
             N^\circ;\mu\rangle
             \longrightarrow^\circ 
             \langle \Gamma'; 
             \overline{\lambda} 
             (x{:}T)^{\ECalpha}. 
             N^{\circ\prime};\mu'\rangle
             }
  \RuleTag{$\xi$-code-$\bar{\lambda}$}
  \quad \cdots
\end{gather*}

\hfill\Fbox{$
  \langle\Gamma;M;\mu\rangle
  \longrightarrow^m
  \langle\Gamma';N;\mu'\rangle
  $}
\begin{gather*}
  \inference{e \vdash \langle\Gamma;M^\circ;\mu\rangle
             \longrightarrow^\circ
             \langle\Gamma';N^\circ;\mu'\rangle}
            {\langle\Gamma;\CodeQuote{M^\circ}e;\mu\rangle
             \longrightarrow^m
             \langle\Gamma';\CodeQuote{N^\circ}e;\mu'\rangle}
  \RuleTag{$\xi$-quote}
  \\[0.5ex]
  \inference{}
            {\langle\Gamma;(e_1{<:}e_2{\Rightarrow}M)\bullet;\mu\rangle
             \longrightarrow^m
             \langle\Gamma;M;\mu\rangle}
  \RuleTag{$\beta$-sub}
  \quad \cdots
\end{gather*}
\caption{Runtime language and selected reduction rules of \LAlpha (excerpt of
  Figures~\ref{fig:runtime-lambda-alpha-ref-types} and~\ref{fig:lambda-alpha-ref-reduction-full}
  of the Appendix).}
\label{fig:lambda-alpha-ref-reduction}
\end{figure}

Figure~\ref{fig:lambda-alpha-ref-reduction} presents selected reduction rules
for \LAlpha. Each configuration consists of a runtime context ($\Gamma$), a
term, and a heap ($\mu$). The heap maps addresses to values. Code and meta
reduction rules are mutually recursive through quotation and splice. We first
define the runtime language, which extends \LAlpha with terms that arise at
runtime. We then describe the reduction semantics. The full version of the
reduction semantics is in Figure~\ref{fig:lambda-alpha-ref-reduction-full} of
the Appendix.

\textbf{Runtime language of \LAlpha.} The runtime language extends \LAlpha with
two terms that arise only at runtime, addresses and code abstraction values
($\bar{\lambda}$). Their syntax and typing rules are defined at the top of
Figure~\ref{fig:lambda-alpha-ref-reduction}. A code abstraction value is a code
abstraction whose classifier and parameter have been allocated in the global
runtime context $\Gamma$. Its typing rule (${\vdash}$\textit{code-abs-val})
checks the allocated entries in $\Gamma$ (highlighted). The global allocation
for code variables allows open code to be stored in the heap while keeping the
heap well typed. An address is typed by the heap typing context $\Sigma$
(${\vdash}$\textit{addr}), which the typing judgments of the runtime language
carry in addition to $\Gamma$.


\textbf{Reducing code terms.} The code reduction $e \vdash
\langle\Gamma;M^\circ;\mu\rangle \longrightarrow^\circ
\langle\Gamma';N^\circ;\mu'\rangle$ reduces a code term under the enclosing
classifier $e$. The reduction (1) eliminates splices and (2) allocates variables
and classifiers of code $\lambda$-abstractions globally. A splice first reduces
its sub-term by the meta reduction ($\xi$\textit{-splice}). The splice then
cancels against the quotation that the sub-term produces
($\beta$\textit{-splice}). The \textit{code-}$\lambda$ rule performs the global
allocation. It first allocates a fresh classifier \ECbeta{} for the scope and a
fresh code variable \CodeInline@y@. It then extends the runtime context $\Gamma$
with \ECbeta{}, the edge $e <: \ECbeta$ to the enclosing scope, and the code
variable entry $(y{:}T)^{\ECbeta}$ (highlighted). Finally, it turns the code
$\lambda$-abstraction into a code abstraction value, renaming variables to
globally allocated ones. A code abstraction value then reduces its body under
its classifier annotation ($\xi$\textit{-code-}$\bar{\lambda}$).

\textbf{Reducing meta terms.} The meta reduction takes the form
$\langle\Gamma;M;\mu\rangle \longrightarrow^m \langle\Gamma';N;\mu'\rangle$. The
reduction follows that of a call-by-value $\lambda$-calculus with references,
except when the term contains a quotation or discharges a subtype constraint. A
quotation reduces the code inside by turning to code reduction
($\xi$\textit{-quote}). A constraint elimination reduces by simply discarding
the subtype constraint, because the subtyping is already checked statically
($\beta$\textit{-sub}).

\section{Target Language: \LCtx{} with Contextual Modal Types}
\label{sec:target-language}

In this section, we define \LCtx, the target language of the compilation. \LCtx
is a two-level calculus based on CMTT~\cite{Nanevski:2008aa}, so a code type
records the context of the variables that the code may mention, and a splice
applies an explicit substitution from one context to another. To these, \LCtx
adds context polymorphism from \LForallBox~\cite{murase2023contextual}, and
general mutable references from Contextual MetaML~\cite{yin2026contextual}. To
support compilation, \LCtx also includes two additional language features:
first-class substitutions and reference proxies. We define the syntax and types
in \S~\ref{sec:lctx-syntax}, the type system in \S~\ref{sec:lctx-typing}, and
the operational semantics in \S~\ref{sec:lctx-semantics}.

\subsection{Types and Syntax of \LCtx}
\label{sec:lctx-syntax}

\begin{figure}[tbp]
\footnotesize
\[
\begin{array}{lrcl}
\text{context variables} & \alpha, \beta, \gamma \\
\text{code typing contexts} & \Gamma^\circ, \Delta^\circ & ::= &
  \emptyset \mid \alpha \mid x{:}T \mid \Gamma^\circ, \Delta^\circ \\
\text{meta types}        & A, B & ::= & \iota \mid A \to B \mid \REFTY{A}
  \mid \CQuote{\Gamma^\circ}{T} \\
  & & \mid & \forall \alpha.\,A \mid \SubstTy{\Gamma^\circ}{\Delta^\circ} \\
\text{meta typing contexts} & \Gamma & ::= & \emptyset \mid \Gamma, \alpha \mid \Gamma, x{:}A \\
\text{code terms}        & L^\circ, M^\circ, N^\circ & ::= &
  k \mid x \mid \lambda (x{:}T).\,N^\circ \mid L^\circ\;M^\circ
  \mid \SPLICE{M}\,N \\
\text{meta terms}        & L, M, N & ::= &
  k \mid x \mid a \mid \lambda (x{:}A).\,N \mid L\;M \\
  & & \mid & \Lambda \alpha.\,N \mid M[\Gamma^\circ] \mid \CQuote{\Gamma^\circ}{M^\circ} \\
  & & \mid & \IdSub{\Gamma^\circ} \mid (M, x{=}N^\circ) \mid \WeakenSub{\Delta^\circ}{M} \mid M \circ N \\
  & & \mid & \REF{M} \mid L := M \mid \mathop{!}M \mid \Proxy{L}{M}{N}
\end{array}
\]
\caption{Syntax of \LCtx.}
\label{fig:lctx-syntax}
\end{figure}

Figure~\ref{fig:lctx-syntax} defines the syntax of \LCtx, where most language
constructs resemble those in \LAlpha. We describe the CMTT-specific features of
\LCtx below.

\textbf{Contexts in types and quotations.} Standard in a CMTT-based calculus, a
quotation $\CQuote{\Gamma^\circ}{M^\circ}$ and a code type
$\CQuote{\Gamma^\circ}{T}$ carry a code context $\Gamma^\circ$. We represent a
code typing context as a binary tree whose leaves are the empty context, a
single mapping $x{:}T$, and context variables $\alpha$. The inner node
$\Gamma^\circ, \Delta^\circ$ concatenates two contexts. We use a binary tree
because this representation is closed under instantiating a context variable.
Replacing the leaf $\alpha$ by a context yields a context, and the rest of the
tree is unchanged.

\textbf{Polymorphism over contexts.} We follow
\LForallBox~\cite{murase2023contextual} and introduce context polymorphism to
\LCtx. The universal type $\forall \alpha.\,A$ quantifies over contexts. Context
abstraction $\Lambda \alpha.\,N$ binds a context variable, and context
application $M[\Gamma^\circ]$ instantiates it with a context.

\textbf{First-class substitutions.} Standard in CMTT, a splice $\SPLICE{M}\,N$
carries a substitution $N$ that maps the variables of the spliced code's context
to code in the current context. Unique to \LCtx, substitutions are first-class
metalanguage terms to support the compilation of bounded quantification
(\S~\ref{sec:examples-polymorphism}).
A substitution term of type
$\SubstTy{\Gamma^\circ}{\Delta^\circ}$ substitutes code in $\Delta^\circ$ for
the variables of $\Gamma^\circ$. Its four introduction forms follow the calculus
of explicit substitutions~\cite{Abadi:1991fr}. The identity
$\IdSub{\Gamma^\circ}$ maps every variable of $\Gamma^\circ$ to itself, the
extension $(M, x{=}N^\circ)$ maps $x$ to $N^\circ$ and the other variables as
$M$ does, the weakening $\WeakenSub{\Delta^\circ}{M}$ lets $M$ target a context
extended by $\Delta^\circ$, and the composition $M \circ N$ applies $M$ and then
$N$.

\textbf{Reference proxies.} Following \citet{Herman:2010aa}, we introduce
proxies of mutable references to \LCtx to support the compilation of subtyping
between reference types. We use the syntax $\Proxy{L}{M}{N}$, which wraps a
reference $L$. The proxy applies the function $M$ to every value read from $L$,
and applies $N$ to every value written to $L$.

\subsection{Type System of \LCtx}
\label{sec:lctx-typing}

\begin{figure}[tbp]
\footnotesize
\hfill\Fbox{$\Gamma; \Sigma; \Gamma^\circ \vdash M^\circ : T$}
\vspace{-5pt}
\begin{gather*}
  \inference{x{:}T \in \Gamma^\circ}
            {\Gamma; \Sigma; \Gamma^\circ \vdash x : T}
  \RuleTag{${\vdash}$code-var}
  \quad
  \inference{\Gamma; \Sigma; (\Gamma^\circ, x{:}S) \vdash N^\circ : T}
            {\Gamma; \Sigma; \Gamma^\circ \vdash \lambda (x{:}S).\,N^\circ : S \to T}
  \RuleTag{${\vdash}$code-abs}
  \\[1ex]
  \inference{\Gamma; \Sigma \vdash M : \CQuote{\Gamma^\circ_1}{T}
             & \Gamma; \Sigma \vdash N : \SubstTy{\Gamma^\circ_1}{\Gamma^\circ_2}}
            {\Gamma; \Sigma; \Gamma^\circ_2 \vdash \SPLICE{M}\,N : T}
  \RuleTag{${\vdash}$splice}
  \qquad
  \cdots
\end{gather*}
\hfill\Fbox{$\Gamma; \Sigma \vdash M : A$}
\vspace{-25pt}
\begin{gather*}
  \inference{\Gamma; \Sigma \vdash L : \REFTY{A} \\
             \Gamma; \Sigma \vdash M : A \to B   \\
             \Gamma; \Sigma \vdash N : B \to A}
            {\Gamma; \Sigma \vdash \Proxy{L}{M}{N} : \REFTY{B}}
  \RuleTagHighlight{${\vdash}$proxy}
  \;
  \inference{\Gamma \vdash \Gamma^\circ \;\mathsf{ctx}
    & \Gamma; \Sigma; \Gamma^\circ \vdash M^\circ : T}
            {\Gamma; \Sigma \vdash \CQuote{\Gamma^\circ}{M^\circ} : \CQuote{\Gamma^\circ}{T}}
  \RuleTag{${\vdash}$quote}
  \\[1ex]
  \inference{\Gamma \vdash \Gamma^\circ \;\mathsf{ctx}}
            {\Gamma; \Sigma \vdash \IdSub{\Gamma^\circ} : \SubstTy{\Gamma^\circ}{\Gamma^\circ}}
  \RuleTagHighlight[GreenYellow!50]{${\vdash}$id}
  \;\;
  \inference{\Gamma \vdash \Gamma^\circ \;\mathsf{ctx}}
            {\Gamma; \Sigma \vdash \WeakenSub{\Gamma^\circ}{\IdSub{\emptyset}} : \SubstTy{\emptyset}{\Gamma^\circ}}
  \RuleTagHighlight[GreenYellow!50]{${\vdash}$weaken-empty}
  \\[1ex]
  \inference{\Gamma; \Sigma \vdash M : \SubstTy{\Gamma^\circ_1}{\Gamma^\circ_2} &
             \Gamma \vdash \Gamma^\circ_3 \;\mathsf{ctx}}
            {\Gamma; \Sigma \vdash \WeakenSub{\Gamma^\circ_3}{M} : \SubstTy{\Gamma^\circ_1}{(\Gamma^\circ_2, \Gamma^\circ_3)}}
  \RuleTagHighlight[GreenYellow!50]{${\vdash}$weaken}
  \\[1ex]
  \inference{\Gamma; \Sigma \vdash M : \SubstTy{\Gamma^\circ}{\Delta^\circ} &
             \Gamma; \Sigma; \Delta^\circ \vdash N^\circ : T}
            {\Gamma; \Sigma \vdash (M, x{=}N^\circ) : \SubstTy{(\Gamma^\circ, x{:}T)}{\Delta^\circ}}
  \RuleTagHighlight[GreenYellow!50]{${\vdash}$ext}
  \\[1ex]
  \inference{\Gamma; \Sigma \vdash M : \SubstTy{\Gamma^\circ_1}{\Gamma^\circ_2} &
             \Gamma; \Sigma \vdash N : \SubstTy{\Gamma^\circ_2}{\Gamma^\circ_3}}
            {\Gamma; \Sigma \vdash M \circ N : \SubstTy{\Gamma^\circ_1}{\Gamma^\circ_3}}
  \RuleTagHighlight[GreenYellow!50]{${\vdash}$compose}
  \qquad \cdots
\end{gather*}
\caption{Selected typing rules of \LCtx{} (excerpt of
  Figure~\ref{fig:lctx-typing-full} of the Appendix).}
\label{fig:lctx-typing}
\end{figure}

Figure~\ref{fig:lctx-typing} shows selected typing rules of \LCtx, and the
complete rules are in Figure~\ref{fig:lctx-typing-full} of the Appendix. All
judgments carry a meta typing context $\Gamma$, which binds context and meta
variables. The term typing judgments also carry a heap typing context $\Sigma$,
whose definition is standard, mapping addresses to types.

\textbf{Well-formed types.} A meta type is well formed, written $\Gamma \vdash A
\;\mathsf{ty}$ , when every type and code context inside it is well formed. A
code context is well formed, written $\Gamma \vdash \Gamma^\circ
\;\mathsf{ctx}$, when all its context variables are well scoped in $\Gamma$.
Their definitions are in Figure~\ref{fig:lctx-wf} of the Appendix.

\textbf{Typing code terms.} The typing judgment for code terms, $\Gamma; \Sigma;
\Gamma^\circ \vdash M^\circ : T$, carries the code context $\Gamma^\circ$. A
code variable is typed by looking the variable up in $\Gamma^\circ$
(${\vdash}$\textit{code-var}). The lookup relation $x{:}T \in \Gamma^\circ$
finds the leaf $(x{:}T)$ in the tree $\Gamma^\circ$. A code abstraction extends
$\Gamma^\circ$ with its parameter for the abstraction's body
(${\vdash}$\textit{code-abs}). The rule for splice (${\vdash}$\textit{splice})
is the one rule that connects two code contexts. Its first argument $M$ produces
code in some context $\Gamma^\circ_1$, and its second argument $N$ is a
substitution from $\Gamma^\circ_1$ to the current context $\Gamma^\circ_2$.

\textbf{Typing meta terms.} The typing judgment for meta terms takes the form
$\Gamma; \Sigma \vdash M : A$. A quotation $\CQuote{\Gamma^\circ}{M^\circ}$ has
type $\CQuote{\Gamma^\circ}{T}$ when $M^\circ$ has type $T$ in $\Gamma^\circ$
(${\vdash}$\textit{quote}). Context abstraction and application follow System~F
with contexts in place of types, and their rules are in the Appendix. Now we
discuss the typing of the new features in \LCtx that support compilation:
reference proxies and first-class substitutions. A proxy of a reference of type
$\REFTY{A}$, with an output function of $A \to B$ and an input function of $B
\to A$, has type $\REFTY{B}$ (${\vdash}$\textit{proxy}, highlighted in blue).
The substitution rules, highlighted in green, type a substitution by its domain
and codomain. The identity on $\Gamma^\circ$ goes from $\Gamma^\circ$ to itself
(${\vdash}$\textit{id}). Weakening the identity on empty gives the empty
substitution into any context (${\vdash}$\textit{weaken-empty}). A weakening
$\WeakenSub{\Gamma^\circ_3}{M}$ keeps the domain of $M$ and joins its codomain
$\Gamma^\circ_2$ with $\Gamma^\circ_3$ (${\vdash}$\textit{weaken}). An extension
$(M, x{=}N^\circ)$ adds $x{:}T$ to the domain of $M$, where $T$ is the type of
$N^\circ$ (${\vdash}$\textit{ext}). A composition $M \circ N$ chains two
substitutions whose contexts match in the middle (${\vdash}$\textit{compose}).

\subsection{Operational Semantics of \LCtx}
\label{sec:lctx-semantics}

\begin{figure}[tbp]
\footnotesize
\setlength{\abovedisplayskip}{2pt}\setlength{\belowdisplayskip}{2pt}%
\[
\begin{array}{lrcl}
\text{code values} & V^\circ, W^\circ & ::= &
  k \mid x \mid \lambda (x{:}T).\,V^\circ \mid V^\circ\;W^\circ \\
\text{meta values} & V, W & ::= &
  k \mid a \mid \lambda (x{:}A).\,N \mid \Lambda \alpha.\,N
  \mid \CQuote{\Gamma^\circ}{V^\circ} \mid \Proxy{V}{V}{V} \\
  & & \mid & \IdSub{\Gamma^\circ} \mid (V, x{=}V^\circ)
  \mid \WeakenSub{\Gamma^\circ}{V} \mid V \circ W \\
\end{array}
\]
\vspace{5pt}
\hfill\Fbox{$M^\circ \mid \mu \longrightarrow^\circ N^\circ \mid \mu'$}
\vspace{-5pt}
\begin{gather*}
  \inference{M \mid \mu \longrightarrow^m M' \mid \mu'}
            {\SPLICE{M}\,N \mid \mu \longrightarrow^\circ \SPLICE{M'}\,N \mid \mu'}
  \RuleTag{$\xi_1$-splice}
  \quad
  \inference{N \mid \mu \longrightarrow^m N' \mid \mu'}
            {\SPLICE{V}\,N \mid \mu \longrightarrow^\circ \SPLICE{V}\,N' \mid \mu'}
  \RuleTag{$\xi_2$-splice}
  \\[1ex]
  \inference{}
            {\SPLICE{\CQuote{\Gamma^\circ}{V^\circ}}\,V \mid \mu
             \longrightarrow^\circ \Sem{V}(V^\circ) \mid \mu}
  \RuleTagHighlight[GreenYellow!50]{$\beta$-splice}
  \qquad
  \cdots
\end{gather*}
\hfill\Fbox{$M \mid \mu \longrightarrow^m N \mid \mu'$}
\vspace{-5pt}
\begin{gather*}
  \inference{M^\circ \mid \mu \longrightarrow^\circ N^\circ \mid \mu'}
            {(M, x{=}M^\circ) \mid \mu \longrightarrow^m (M, x{=}N^\circ) \mid \mu'}
  \RuleTag{$\xi$-ext}
  \\[0.7ex]
  \inference{}
            {(\Lambda \alpha.\,N)[\Gamma^\circ] \mid \mu \longrightarrow^m N[\alpha := \Gamma^\circ] \mid \mu}
  \RuleTag{$\beta$-ctx}
  \\[0.7ex]
  \inference{}
            {\mathop{!}(\Proxy{V_1}{V_2}{V_3}) \mid \mu \longrightarrow^m V_2\;(\mathop{!}V_1) \mid \mu}
  \RuleTag{proxy-deref}
  \\[0.7ex]
  \inference{}
            {(\Proxy{V_1}{V_2}{V_3}) := W \mid \mu \longrightarrow^m V_1 := (V_3\;W) \mid \mu}
  \RuleTag{proxy-assign}
  \qquad\cdots
\end{gather*}
\caption{Selected reduction rules of \LCtx{} (excerpt of
  Figure~\ref{fig:lctx-semantics-full} of the Appendix).}
\label{fig:lctx-semantics}
\end{figure}

Figure~\ref{fig:lctx-semantics} shows selected reduction rules of \LCtx, and the
complete rules are in Figure~\ref{fig:lctx-semantics-full} of the Appendix. The
semantics consists of two mutually recursive reduction relations, one for code
and the other for meta.

\textbf{Values.} A code value is a term in STLC with no remaining splice. The
meta values are constants, addresses, abstractions, quotations of code values,
proxies of values, and substitution values. A substitution value is built from
the four substitution introduction forms, each with values in its sub-terms. Its
meaning is given by the interpretation $\Sem{V}$, defined in
Figure~\ref{fig:lctx-semantics-full} of the Appendix. The interpretation of $V$
is a substitution that substitutes code for each variable in the domain of $V$.
Applying $\Sem{V}$ to a code value performs the substitution.

\textbf{Reducing code terms.} The code reduction takes the form $M^\circ \mid
\mu \longrightarrow^\circ N^\circ \mid \mu'$. Code reduces only to eliminate
splices. A splice first reduces its meta term to a quotation
($\xi_1$\textit{-splice}) and its substitution to a substitution value
($\xi_2$\textit{-splice}). The rule $\beta$\textit{-splice} is the heart of the
semantics, highlighted in green. It reduces the splice of a quotation
$\CQuote{\Gamma^\circ}{V^\circ}$ under a substitution value $V$ to
$\Sem{V}(V^\circ)$, which is the quoted code with the substitution applied.

\textbf{Reducing meta terms.} The meta reduction takes the form $M \mid \mu
\longrightarrow^m N \mid \mu'$. Meta terms reduce as in a call-by-value
$\lambda$-calculus with references, except in three cases. The code attached to
a substitution extension reduces when it contains splices ($\xi$\textit{-ext}).
Context application substitutes the context $\Gamma^\circ$ for the context
variable $\alpha$ in the body ($\beta$\textit{-ctx}). A proxy forwards
dereference and assignment to the reference it wraps, applying its output
function to every value read (\textit{proxy-deref}) and its input function to
every value written (\textit{proxy-assign}).

\section{Compilation From \LAlpha{} to \LCtx}
\label{sec:compilation}

In this section, we formally define the compilation from \LAlpha to \LCtx. The
compilation has three parts. First, classifiers compile to code contexts, and
types compile accordingly (\S~\ref{sec:compile-types}). Second, subtyping
between classifiers compiles to substitutions, and subtyping between types
compiles to coercion functions (\S~\ref{sec:compile-subtyping}). Third, terms
compile according to their typing derivations, with classifier polymorphism
becoming context polymorphism and subtype constraints becoming abstraction over
and application to substitutions (\S~\ref{sec:compile-terms}).

\subsection{Compiling Classifiers and Types}
\label{sec:compile-types}

\begin{figure}[tbp]
\footnotesize
\[
\begin{array}{lrcl}
\text{compilation contexts} & \Xi & ::= & \emptyset \mid \Xi, \ECalpha
  \mid \Xi, \RuleHighlight[GreenYellow!50]{\ECalpha_{\mathsf{poly}}} \mid \Xi, x{:}A
  \mid \Xi, \RuleHighlight[GreenYellow!50]{s : e_1 \sub e_2} \\
  & & \mid & \Xi, (x{:}T)^{\ECalpha} \mid \Xi, e_1 \sub e_2
\end{array}
\]
\hfill\Fbox{$\Xi \vdash e \compile \Gamma^\circ$}
\vspace{-15pt}
\begin{gather*}
  \inference{}{\Xi \vdash \ECeps \compile \emptyset}
  \RuleTag{${\compile}$empty}
  \;
  \inference{\ECalpha \in \Xi & e \sub \ECalpha \in \Xi \\ (x{:}T)^{\ECalpha} \in \Xi
    & \Xi \vdash e \compile \Gamma^\circ}
            {\Xi \vdash \ECalpha \compile (\Gamma^\circ, x{:}T)}
            \RuleTag{${\compile}$ec}
  \;
  \inference{\ECalpha_{\mathsf{poly}} \in \Xi}
            {\Xi \vdash \ECalpha \compile \alpha}
  \RuleTag{${\compile}$poly-ec}
\end{gather*}
\hfill\Fbox{$\Xi \vdash A \compile A'$}
\vspace{-25pt}
\begin{gather*}
  \inference{\Xi \vdash e \compile \Gamma^\circ}
            {\Xi \vdash \Quote{T}{e} \compile \CQuote{\Gamma^\circ}{T}}
  \RuleTag{${\compile}$code}
  \\[0.7ex]
  \inference{\Xi \vdash e_1 \sub e_2 \compile B &
             \Xi, s : e_1 \sub e_2 \vdash A \compile A' & s \notin \Xi}
            {\Xi \vdash (e_1 \sub e_2 \Rightarrow A) \compile (B \to A')}
  \RuleTag{${\compile}$constraint}
  \qquad \cdots
\end{gather*}
\caption{Compilation contexts, and selected compilation rules of
  classifiers and types (excerpt of Figure~\ref{fig:compile-types-full} of
  the Appendix).}
\label{fig:compile-types}
\end{figure}

Figure~\ref{fig:compile-types} shows selected compilation rules for classifiers
and types. The complete compilation rules are in
Figure~\ref{fig:compile-types-full} of the Appendix.

\textbf{Compilation contexts.} A compilation context $\Xi$, defined at the top
of Figure~\ref{fig:compile-types}, is a typing context of \LAlpha extended by
the two highlighted entries. First, a compilation context distinguishes the two
ways a classifier enters scope, because they compile differently. A classifier
bound by a code abstraction is recorded as $\ECalpha$ and compiles to a concrete
context, whereas a classifier bound by a classifier abstraction is recorded as
$\ECalpha_{\mathsf{poly}}$ and compiles to a context variable. Second, it
distinguishes the two kinds of subtype constraints. A constraint $e_1 \sub e_2$
comes from a code abstraction, whereas a named constraint $s : e_1 \sub e_2$
comes from a constraint introduction and binds the variable $s$ as evidence for
the constraint.

\textbf{Compiling classifiers to contexts.} The judgment $\Xi \vdash e \compile
\Gamma^\circ$ computes the context that corresponds to a classifier. The empty
classifier compiles to the empty context (${\compile}$\textit{empty}), and a
polymorphic classifier compiles to a context variable
(${\compile}$\textit{poly-ec}). The interesting case is a classifier $\ECalpha$
introduced by a code abstraction (${\compile}$\textit{ec}). Its context is
calculated by following the subtyping in $\Xi$. We first
include the code variable entry $(x{:}T)^{\ECalpha}$ in the result context. We then follow the
edge $e \sub \ECalpha$, which leads to the enclosing classifier $e$. The context
of $e$ is computed in turn. Following the edges up to $\ECeps$ or to a
polymorphic classifier, we collect exactly the variables that the scope of
$\ECalpha$ may mention.

\textbf{Compiling types.} The judgment $\Xi \vdash A \compile A'$ compiles
types, and its rules mirror the well-formedness rules of types in \LAlpha. A
code type $\Quote{T}{e}$ compiles to $\CQuote{\Gamma^\circ}{T}$, where
$\Gamma^\circ$ is the context of $e$ (${\compile}$\textit{code}). A
subtype-constrained type $e_1 \sub e_2 \Rightarrow A$ compiles to the function
type $B \to A'$ (${\compile}$\textit{constraint}). Its argument type $B$ is the
type of the evidence for the constraint, computed by the judgment $\Xi \vdash
e_1 \sub e_2 \compile B$ that we define in the next subsection. Its body $A$ is
compiled to $A'$ under $\Xi$ extended with the evidence variable $s$ for the
constraint. The remaining types compile recursively.

\begin{figure}[tbp]
\footnotesize
\hfill\Fbox{$\Xi \vdash e_1 \sub e_2 \compile A$}
\vspace{-15pt}
\begin{gather*}
  \inference{\Xi \vdash e_1 \compile \Gamma^\circ_1 & \Xi \vdash e_2 \compile \Gamma^\circ_2}
            {\Xi \vdash e_1 \sub e_2 \compile \SubstTy{\Gamma^\circ_1}{\Gamma^\circ_2}}
  \RuleTag{${\compile}$sub-ec}
\end{gather*}
\hfill\Fbox{$\Xi \vdash e_1 \sub e_2 \compile M$}
\begin{gather*}
  \inference{\Xi \vdash e \compile \Gamma^\circ}
            {\Xi \vdash \ECeps \sub e \compile \WeakenSub{\Gamma^\circ}{\IdSub{\emptyset}}}
  \RuleTag{${\compile}$sub-empty}
  \quad
  \inference{\Xi \vdash e \compile \Gamma^\circ}
            {\Xi \vdash e \sub e \compile \IdSub{\Gamma^\circ}}
  \RuleTag{${\compile}$sub-refl}
  \\[1ex]
  \inference{e \sub \ECalpha \in \Xi & (x{:}T)^{\ECalpha} \in \Xi \\ \Xi \vdash e \compile \Gamma^\circ}
            {\Xi \vdash e \sub \ECalpha \compile \WeakenSub{x{:}T}{\IdSub{\Gamma^\circ}}}
  \RuleTag{${\compile}$sub-ax}
  \quad
  \inference{(s : e_1 \sub e_2) \in \Xi}
            {\Xi \vdash e_1 \sub e_2 \compile s}
  \RuleTag{${\compile}$sub-var}
  \\[1ex]
  \inference{\Xi \vdash e_1 \sub e_3 \compile M & \Xi \vdash e_3 \sub e_2 \compile N}
            {\Xi \vdash e_1 \sub e_2 \compile M \circ N}
  \RuleTag{${\compile}$sub-trans}
\end{gather*}
\hfill\Fbox{$\Xi \vdash A \sub B \compile M$}
\vspace{-5pt}
\begin{gather*}
  \inference{\Xi \vdash e_1 \compile \Gamma^\circ & \Xi \vdash e_2 \compile \Delta^\circ
             & \RuleHighlight[GreenYellow!50]{\Xi \vdash e_1 \sub e_2 \compile M}}
            {\Xi \vdash \Quote{T}{e_1} \sub \Quote{T}{e_2} \compile
             \lambda (c{:}\CQuote{\Gamma^\circ}{T}).\,\CQuote{\Delta^\circ}{\SPLICE{c}\,\RuleHighlight[GreenYellow!50]{M}}}
  \RuleTag{${\compile}$sub-code}
  \\[1ex]
  \inference{\Xi \vdash A \compile A' & \Xi \vdash A \sub B \compile M & \Xi \vdash B \sub A \compile N}
            {\Xi \vdash \REFTY{A} \sub \REFTY{B} \compile \lambda (r{:}\REFTY{A'}).\,\Proxy{r}{M}{N}}
  \RuleTag{${\compile}$sub-ref}
  \\[1ex]
  \inference{\Xi \vdash e_1 \sub e_2 \compile C
             & \Xi, (s : e_1 \sub e_2) \vdash A \compile A' \\
             \Xi, (s : e_1 \sub e_2) \vdash A \sub B \compile M & s \notin \Xi}
            {\begin{array}[t]{@{}l@{}}
             \Xi \vdash (e_1 \sub e_2 \Rightarrow A) \sub (e_1 \sub e_2 \Rightarrow B) \compile {} \\
             \qquad\qquad \lambda (f{:}C \to A').\,\lambda (s{:}C).\,M\;(f\;s)
             \end{array}}
  \RuleTag{${\compile}$sub-constraint}
  \qquad \cdots
\end{gather*}
\caption{Compilation of subtyping between classifiers and between types \\
  (excerpt of Figure~\ref{fig:compile-subtyping-full} of the Appendix).}
\label{fig:compile-subtyping}
\end{figure}

\subsection{Compiling Subtyping}
\label{sec:compile-subtyping}

Figure~\ref{fig:compile-subtyping} presents selected compilation rules of
subtyping. The complete rules are in Figure~\ref{fig:compile-subtyping-full} of
the Appendix. The compilation follows the coercion semantics of
subtyping~\cite{Breazu-Tannen:1991aa,Pierce:2002hj}. Subtyping between
classifiers compiles to a first-class substitution, and subtyping between types
compiles to a coercion function.

\textbf{Compiling classifier subtyping to a substitution.} The judgment $\Xi
\vdash e_1 \sub e_2 \compile M$ compiles two classifiers $e_1$ and $e_2$, the
former a subtype of the latter, to a substitution $M$. The type of $M$ is given
by $\Xi \vdash e_1 \sub e_2 \compile A$, which produces a substitution type from
the compiled context of $e_1$ to that of $e_2$. The rules of $\Xi \vdash e_1
\sub e_2 \compile M$ mirror the subtyping rules for classifiers in \LAlpha. The
axiom $\ECeps \sub e$ compiles to the empty substitution into the context of $e$
(${\compile}$\textit{sub-empty}). Reflexivity compiles to the identity
(${\compile}$\textit{sub-refl}). The two kinds of subtyping entries in $\Xi$
compile differently. A subtyping entry $e \sub \ECalpha$, introduced by a code
$\lambda$, compiles to the identity on the context of $e$, weakened by the code
variable $(x{:}T)$ bound by that code $\lambda$ (${\compile}$\textit{sub-ax}). A
named constraint, introduced by a subtype constraint, compiles to the variable
$s$ for the evidence of that constraint (${\compile}$\textit{sub-var}).
Transitivity compiles to composition (${\compile}$\textit{sub-trans}).

\textbf{Compiling type subtyping to a coercion.} The judgment $\Xi \vdash A \sub
B \compile M$ compiles two types $A$ and $B$, with $A$ a subtype of $B$, to a
coercion function $M$. The compilation rules mirror the subtyping rules for
types in \LAlpha. The coercion converts from the compiled type of $A$ to that of
$B$. It applies the substitution that the underlying classifier subtyping
compiles to, and the rule for code types is the key
(${\compile}$\textit{sub-code}). The subtyping $\Quote{T}{e_1} \sub
\Quote{T}{e_2}$ compiles to a function that takes code in the context that $e_1$
corresponds to, $\Gamma^\circ$, and moves it to the context that $e_2$
corresponds to, $\Delta^\circ$. The body of the function splices the code under
the substitution that $e_1 \sub e_2$ compiles to (highlighted) and re-quotes the
result in $\Delta^\circ$. A reference type is special, because a reference can
be both read from and written to. Consequently, its coercion wraps the reference
in a proxy whose output and input functions are the coercions in the two
directions, read and write (${\compile}$\textit{sub-ref}). A subtype-constrained
type also deserves a closer look (${\compile}$\textit{sub-constraint}). The
types $e_1 \sub e_2 \Rightarrow A$ and $e_1 \sub e_2 \Rightarrow B$ compile to
the function types $C \to A'$ and $C \to B'$, where $C$ is the type of the
evidence for the constraint, and $A', B'$ are what $A, B$ compile to,
respectively. The coercion is thus a function between function types, and the
compilation follows the types. The coercion takes a function $f$ and evidence
$s$, applies $f$ to $s$, and coerces the result. The remaining cases follow the
standard coercion semantics.

\subsection{Compiling Terms}
\label{sec:compile-terms}

\begin{figure}[tbp]
\footnotesize
\hfill\Fbox{$\Xi; e \vdash M^\circ : T \compile M^{\circ\prime}$}
\begin{gather*}
  \inference{\Xi, \ECalpha, e \sub \ECalpha, (x{:}S)^{\ECalpha}; \ECalpha \vdash N^\circ : T \compile N^{\circ\prime}
             & \ECalpha \notin \Xi}
            {\Xi; e \vdash \lambda (x{:}S)^{\ECalpha}.\,N^\circ : S \to T \compile \lambda (x{:}S).\,N^{\circ\prime}}
  \RuleTag{${\compile}$code-abs}
  \\[1ex]
  \inference{\Xi \vdash M : \Quote{T}{e} \compile M' &
             \Xi \vdash e \compile \Gamma^\circ}
            {\Xi; e \vdash \SPLICE{M} : T \compile \SPLICE{M'}\,\IdSub{\Gamma^\circ}}
  \RuleTag{${\compile}$splice}
  \\[1ex]
  \inference{\Xi; e_1 \vdash M^\circ : T \compile M^{\circ\prime}
             & \Xi \vdash e_1 \compile \Gamma^\circ
             & \RuleHighlight{\Xi \vdash e_1 \sub e_2 \compile M}}
            {\Xi; e_2 \vdash M^\circ : T \compile \SPLICE{\CQuote{\Gamma^\circ}{M^{\circ\prime}}}\,\RuleHighlight{M}}
  \RuleTag{${\compile}$subenv}
  \quad \cdots
\end{gather*}
\hfill\Fbox{$\Xi \vdash M : A \compile M'$}
\vspace{-5pt}
\begin{gather*}
  \inference{\Xi; e \vdash M^\circ : T \compile M^{\circ\prime} &
             \Xi \vdash \RuleHighlight[GreenYellow!50]{e \compile \Gamma^\circ}}
            {\Xi \vdash \Quote{M^\circ}{{\RuleHighlight[GreenYellow!50]{e}}} : \Quote{T}{e} \compile
             \CQuote{\RuleHighlight[GreenYellow!50]{\Gamma^\circ}}{M^{\circ\prime}}}
  \RuleTag{${\compile}$quote}
  \\[1ex]
  \inference{\Xi, \ECalpha_{\mathsf{poly}} \vdash N : A \compile N' & \ECalpha \notin \Xi}
            {\Xi \vdash \Lambda \RuleHighlight[GreenYellow!50]{\ECalpha}.\,N : \forall \ECalpha.\,A \compile \Lambda \RuleHighlight[GreenYellow!50]{\alpha}.\,N'}
  \RuleTag{${\compile}$ec-abs}
  \\[1ex]
  \inference{\Xi \vdash M : \forall \ECalpha.\,A \compile M'
             & \Xi \vdash \RuleHighlight[GreenYellow!50]{e \compile \Gamma^\circ}}
            {\Xi \vdash M[\RuleHighlight[GreenYellow!50]{e}] : A[\ECalpha := e] \compile
             M'[\RuleHighlight[GreenYellow!50]{\Gamma^\circ}]}
  \RuleTag{${\compile}$ec-app}
  \\[1ex]
  \inference{\Xi \vdash e_1 \compile \Gamma^\circ_1 & \Xi \vdash e_2 \compile \Gamma^\circ_2 \\
             \Xi, \RuleHighlight{(s : e_1 \sub e_2)} \vdash N : A \compile N' & s \notin \Xi}
            {\Xi \vdash (e_1 \sub e_2 \Rightarrow N) : (e_1 \sub e_2 \Rightarrow A) \compile
             \lambda (\RuleHighlight{s{:}\SubstTy{\Gamma^\circ_1}{\Gamma^\circ_2}}).\,N'}
  \RuleTag{${\compile}$sub-intro}
  \\[1ex]
  \inference{\Xi \vdash M : (e_1 \sub e_2 \Rightarrow A) \compile M' \\ \RuleHighlight{\Xi \vdash e_1 \sub e_2 \compile N}}
            {\Xi \vdash M\bullet : A \compile M'\;N}
  \RuleTag{${\compile}$sub-elim}
  \;
  \inference{\Xi \vdash M : A \compile M' \\ \RuleHighlight{\Xi \vdash A \sub B \compile L}}
            {\Xi \vdash M : B \compile L\;M'}
  \RuleTag{${\compile}$subsumption}
  \\
  \cdots
\end{gather*}
\caption{Selected compilation rules of terms
  (excerpt of Figure~\ref{fig:compile-terms-full} of the Appendix).}
\label{fig:compile-terms}
\end{figure}

Figure~\ref{fig:compile-terms} shows selected compilation rules for code terms
and meta terms. The complete compilation rules are in
Figure~\ref{fig:compile-terms-full} of the Appendix. The rules mirror the typing
rules of \LAlpha and do two things. First, wherever a typing rule of \LAlpha
mentions a classifier, the compilation rule converts it into the context that
the classifier corresponds to (the green highlights). Second, wherever a typing
rule of \LAlpha uses subtyping, the compiled term carries the substitution or
coercion that the subtyping compiles to (the blue highlights). The remaining
rules compile recursively, except that a code abstraction drops its classifier
(${\compile}$\textit{code-abs}) and a splice gains the identity substitution
that a splice of \LCtx requires (${\compile}$\textit{splice}).

\textbf{Converting classifiers into contexts.} The three rules with green
highlights replace a classifier by its context. The context is computed by $\Xi
\vdash e \compile \Gamma^\circ$ of Figure~\ref{fig:compile-types}. A quotation
$\Quote{M^\circ}{e}$ compiles to $\CQuote{\Gamma^\circ}{M^{\circ\prime}}$
(${\compile}$\textit{quote}). A classifier abstraction compiles to a context
abstraction (${\compile}$\textit{ec-abs}). A classifier application $M[e]$
compiles to a context application $M'[\Gamma^\circ]$
(${\compile}$\textit{ec-app}).

\textbf{Compiling subtyping.} The four rules with blue highlights insert the
substitution or coercion that a subtyping compiles to. Classifier subsumption
from $e_1$ to $e_2$ quotes the code in the context of $e_1$ and splices it under
the substitution for $e_1 \sub e_2$, which carries the code to the context of
$e_2$ (${\compile}$\textit{subenv}). Subtype constraint introduction compiles to
an abstraction over the evidence variable $s$ for the constraint, which has a
substitution type, with the body compiled under $\Xi$ extended with the named
constraint $s : e_1 \sub e_2$ (${\compile}$\textit{sub-intro}). Constraint
elimination $M\bullet$ compiles to an application to the substitution for the
discharged constraint (${\compile}$\textit{sub-elim}). Subsumption from $A$ to
$B$ compiles to an application of the coercion for $A \sub B$
(${\compile}$\textit{subsumption}). In summary, these four rules replace every
use of subtyping in the \LAlpha typing derivation by an explicit term in \LCtx.

\section{Metatheoretical Results}
\label{sec:metatheory}

In this section, we prove that the compilation of Section~\ref{sec:compilation}
preserves types (\S~\ref{sec:type-preservation}) and is semantically correct
(\S~\ref{sec:semantic-correctness}). All lemmas and theorems in this section are
mechanized in Agda. Table~\ref{tab:paper-agda-result-map} of the Appendix maps
each result to Agda.

\subsection{Compilation Preserves Types}
\label{sec:type-preservation}

We first relate compilation contexts to typing contexts. We then define the
well-formedness of compilation contexts. With the definitions in hand, we prove
that the compilation from \LAlpha to \LCtx preserves types
(Theorem~\ref{thm:compile-pres-types}).

\textbf{Relating the contexts.} The lemmas that lead to type preservation
involve three judgments: the typing judgment of \LAlpha, the compilation
judgment $\Xi \vdash M : A \compile M'$, and the typing judgment of \LCtx. Since each
judgment has its own context, $\Xi$ needs a reading in both languages. For the
source language \LAlpha, the operator $\Erase{\Xi}$, defined in
Figure~\ref{fig:compile-ctx} of the Appendix, turns $\Xi$ into a \LAlpha typing
context. It forgets the two annotations that only the compilation needs, namely
which classifiers are polymorphic and which subtype constraints have evidence
variables. For the target language \LCtx, the judgment $\vdash \Xi \compile
\Gamma$, also defined in Figure~\ref{fig:compile-ctx} of the Appendix, compiles
the entries of the metalanguage in $\Xi$ to a meta typing context $\Gamma$.
The code-level entries of $\Xi$ are dropped, because the code context of a code
term comes instead from its classifier.

\textbf{Well-formed compilation contexts.} The compilation lemmas require
compilation contexts to be well formed. Intuitively, a compilation context is
well formed when each scope has at most one parent scope, which may be empty
(\ECeps), and two distinct scopes do not share a classifier.
Definition~\ref{def:wf-ctx} of the Appendix formalizes this intuition, and every
compilation context that the compilation in Section~\ref{sec:compilation} builds
is well formed.

\textbf{Compilation preserves types.}
The first lemma shows that every well-formed classifier compiles to exactly one
context, and every well-formed type to exactly one type, both of which are well
formed in the target context.

\begin{lemma}[Compiled contexts and types are well formed]
\label{lem:ec-unique}
\tightlist
\\ Suppose $\vdash \Xi$ and $\vdash \Xi \compile \Gamma$.
\begin{enumerate}
\item If $\Erase{\Xi} \vdash e$, then $\Xi \vdash e \compile \Gamma^\circ$ and $\Gamma \vdash
  \Gamma^\circ \;\mathsf{ctx}$
  for a unique $\Gamma^\circ$.
\item If $\Erase{\Xi} \vdash A$, then 
  $\Xi \vdash A \compile A'$ and $\Gamma \vdash A' \;\mathsf{ty}$
  for a unique $A'$.
\end{enumerate}
\end{lemma}

The second lemma shows that the terms the compilation inserts for subtyping are
well typed under the meta typing context $\Gamma$ that $\Xi$ compiles to. The
heap typing context is empty because the compilation does not produce addresses.

\begin{lemma}[Subtyping compiles to well-typed terms]
\label{lem:sub-pres}
\tightlist
\\ Suppose $\vdash \Xi$ and $\vdash \Xi \compile \Gamma$.
\begin{enumerate}
\item If $\Xi \vdash e_1 \compile \Gamma^\circ_1$, $\Xi \vdash e_2 \compile
  \Gamma^\circ_2$, and $\Xi \vdash e_1 \sub e_2 \compile M$, \\ then $\Gamma;
  \emptyset \vdash M : \SubstTy{\Gamma^\circ_1}{\Gamma^\circ_2}$.
\item If $\Xi \vdash A \compile A'$, $\Xi \vdash B \compile B'$, and $\Xi \vdash
  A \sub B \compile M$, \\ then $\Gamma; \emptyset \vdash M : A' \to B'$.
\end{enumerate}
\end{lemma}

The main lemma shows that every well-typed \LAlpha term compiles, and that the
compiled term is well typed in \LCtx at the compiled type.

\begin{lemma}[Compilation of code and meta terms preserves types]
\label{lem:compile-pres}
\tightlist
\\ Suppose $\vdash \Xi$ and $\vdash \Xi \compile \Gamma$.
\begin{enumerate}
\item If $\Erase{\Xi}; e \vdash M^\circ : T$ and $\Xi \vdash e \compile
  \Gamma^\circ$, then there exists an $M^{\circ\prime}$ such that $\Xi; e \vdash
  M^\circ : T \compile M^{\circ\prime}$, and $\Gamma; \emptyset; \Gamma^\circ
  \vdash M^{\circ\prime} : T$.
\item If $\Erase{\Xi} \vdash M : A$, then there exist $A'$ and $M'$ such that
  $\Xi \vdash A \compile A'$ and $\Xi \vdash M : A \compile M'$, and $\Gamma;
  \emptyset \vdash M' : A'$.
\end{enumerate}
\end{lemma}

The type preservation theorem is a corollary of Lemma~\ref{lem:compile-pres} for
closed programs.

\begin{theorem}[Compilation of \LAlpha programs to \LCtx preserves types]
  \label{thm:compile-pres-types}
  \setlength{\abovedisplayskip}{3pt}\setlength{\belowdisplayskip}{3pt}%
  \setlength{\abovedisplayshortskip}{0pt}\setlength{\belowdisplayshortskip}{3pt}%
\\
If a \LAlpha term $M$ is well typed, $\emptyset \vdash M : A$, \\
then there exist a type $A'$ and a term $M'$ in \LCtx such that \\
$\emptyset \vdash A \compile A'$,
$\emptyset \vdash M : A \compile M'$, and
$\emptyset; \emptyset \vdash M' : A'$.
\end{theorem}

\subsection{Semantic Correctness of Compilation Through Bisimulation}
\label{sec:semantic-correctness}

We now prove the semantic correctness of compilation
(Theorem~\ref{thm:semantic-correctness}). We first define reification, which
turns code values of \LAlpha into STLC terms. We then prove bisimulation
between \LAlpha and \LCtx and obtain the theorem as a corollary.

\textbf{Reification.} The code value that a \LAlpha program generates contains
globally allocated variables, because the \textit{code-}$\lambda$ rule allocates
a global variable for each code abstraction. To obtain code in STLC, reification
converts each global variable to a fresh lexical one, so that every code
abstraction value ($\overline{\lambda}$) becomes a lexical $\lambda$ of STLC.
Reification is defined in Figure~\ref{fig:reify} of the Appendix. The code
values of \LCtx do not need reification, because they are already STLC terms.

\textbf{Bisimulation.} The simulation relation on terms is similar to the
compilation relation of Section~\ref{sec:compilation}, but it is closed under
reduction. It relates a code or meta term of \LAlpha to one of \LCtx under a
compilation context $\Xi$ and a heap typing context $\Sigma$. The full
simulation rules are in Figures~\ref{fig:sim-code} and~\ref{fig:sim-meta} of the
Appendix. The simulation relation differs from the compilation in two ways.
First, it relates terms that arise only at runtime, such as code abstraction
values ($\overline{\lambda}$) and addresses. Second, it has value rules for
catching up with a coercion. Once a subsumed source term is a value, the coerced
target term still has to apply its coercion or substitution, in steps that the
source does not take. The value simulation rules relate the source value to the
value that the target reaches once it has caught up.

The embedding lemma says that a term is in simulation with its compilation.

\begin{lemma}[Compilation embeds into simulation]
\label{lem:sim-embed}
\tightlist
\\ Let $\Sigma$ be any heap typing context.
\begin{enumerate}
\item If $\Xi; e \vdash M^\circ : T \compile M^{\circ\prime}$, then
  $\Sigma; \Xi; e \vdash M^\circ : T \simrel M^{\circ\prime}$.
\item If $\Xi \vdash M : A \compile M'$, then $\Sigma; \Xi \vdash M : A \simrel M'$.
\end{enumerate}
\end{lemma}

The simulation relation on configurations, written ${\Sigma; \Psi \vdash C : A
  \simrel M \mid \mu}$, relates a configuration $C$ of \LAlpha to a term $M$ and
a heap $\mu$ of \LCtx. The simulation context $\Psi$ holds the classifiers,
subtype constraints, and code variables that code abstractions add: $\Psi ::=
\emptyset \mid \Psi, \ECalpha \mid \Psi, (x{:}T)^{\ECalpha} \mid \Psi, e_1 \sub
e_2$. We write $\longrightarrow^{m*}$ for the reflexive, transitive closure of
a reduction relation and $\subseteq$ for inclusion of contexts viewed as sets.
The next two lemmas establish a bisimulation.

\begin{lemma}[Forward simulation]
\label{lem:forward-sim}
\\ If $\Sigma; \Psi \vdash C : A \simrel M \mid \mu$ and $C
\longrightarrow^m D$, \\
then there exist $\Psi'$, $\Sigma'$, $N$, and $\mu'$ such that $\Psi \subseteq
\Psi'$, $\Sigma \subseteq \Sigma'$, $M \mid \mu \longrightarrow^{m*} N \mid
\mu'$, and $\Sigma'; \Psi' \vdash D : A \simrel N \mid \mu'$.
\end{lemma}

\begin{lemma}[Backward simulation]
\label{lem:backward-sim}
\\ If $\Sigma; \Psi \vdash C : A \simrel M \mid \mu$ and $M \mid \mu
\longrightarrow^m N \mid \mu'$, \\
then there exist $\Psi'$, $\Sigma'$, $D$, $N'$, and $\mu''$ such that $\Psi
\subseteq \Psi'$, $\Sigma \subseteq \Sigma'$, $C \longrightarrow^{m*} D$,
${N \mid \mu' \longrightarrow^{m*} N' \mid \mu''}$, and $\Sigma'; \Psi' \vdash
D : A \simrel N' \mid \mu''$.
\end{lemma}

\textbf{Semantic correctness.} A configuration $C$ of \LAlpha diverges, written
$C \Uparrow$, if every configuration that $C$ reduces to can take a step, and
likewise $M \mid \mu \Uparrow$ in \LCtx. The semantic correctness theorem states
that a closed \LAlpha program $M$ of code type $\Quote{T}{\ECeps}$ and its
compilation $M'$ behave the same. If one of them evaluates to a quotation of a
code value, so does the other, and the code value of $M$ reifies to the
generated code of $M'$. The generated code is an STLC term of type $T$. If one
of them diverges, so does the other. The proof relates $M$ to $M'$ by
Lemma~\ref{lem:sim-embed} and keeps them related by Lemmas~\ref{lem:forward-sim}
and~\ref{lem:backward-sim} as either side reduces.

\begin{theorem}[Semantic correctness]
\label{thm:semantic-correctness}
\tightlist
\setlength{\abovedisplayskip}{3pt}\setlength{\belowdisplayskip}{3pt}%
\setlength{\abovedisplayshortskip}{0pt}\setlength{\belowdisplayshortskip}{3pt}%
\\ Suppose a \LAlpha term $M$ is well typed, $\emptyset \vdash M : \Quote{T}{\ECeps}$. \\
Then there exists a well-typed \LCtx term $M'$ that $M$ compiles to, \\
$\emptyset \vdash M : \Quote{T}{\ECeps} \compile M'$ and
$\emptyset; \emptyset \vdash M' : \CQuote{\emptyset}{T}$,
such that the following hold.
\begin{enumerate}
\item If $\langle\emptyset; M; \emptyset\rangle \longrightarrow^{m*}
  \langle\Gamma; V; \mu\rangle$ for a value $V$, then $V = \Quote{V^\circ}{\ECeps}$
  for a code value $V^\circ$, and there exist a code value $W^\circ$ and a heap
  $\mu'$ such that $M' \mid \emptyset \longrightarrow^{m*}
  \CQuote{\emptyset}{W^\circ} \mid \mu'$ and $\Reify{V^\circ} = W^\circ$.
\item If $M' \mid \emptyset \longrightarrow^{m*} \CQuote{\emptyset}{W^\circ}
  \mid \mu'$ for a code value $W^\circ$, then there exist a code value
  $V^\circ$, a context $\Gamma$, and a heap $\mu$ such that
  $\langle\emptyset; M; \emptyset\rangle \longrightarrow^{m*} \langle\Gamma;
  \Quote{V^\circ}{\ECeps}; \mu\rangle$ and $\Reify{V^\circ} = W^\circ$.
\item $\langle\emptyset; M; \emptyset\rangle \Uparrow$ if and only if
  $M' \mid \emptyset \Uparrow$.
\end{enumerate}
In the first two cases, the generated code is a well-typed STLC term,
${\emptyset \vdash W^\circ : T}$.
\end{theorem}

\section{Conclusion}
\label{sec:conclusion}

This paper connects two approaches to the scopes of generated code,
environment classifiers and contextual modal type theory, through a
type-directed compilation. We proved in Agda that the compilation
preserves types and behavior. CMTT is thus at least as expressive as
environment classifiers. We conjecture that the converse compilation
is hard, because environment classifiers impose a stricter regime on
how the metaprogram scopes generated code.

%
%
%
\bibliographystyle{splncs04nat}
\bibliography{all}

\newpage
\section*{Appendix}
\setcounter{secnumdepth}{0}  

The appendix is organized as follows.
\begin{enumerate}
\item Table~\ref{tab:paper-agda-result-map} maps the metatheoretical results to
  their Agda mechanization.
\item Figure~\ref{fig:lambda-alpha-ref-well-formedness} defines the
  well-formedness for ECs and meta types of \LAlpha{}.
\item Figure~\ref{fig:lambda-alpha-ref-subtyping-full} defines the subtyping
  rules of \LAlpha{}.
\item Figure~\ref{fig:lambda-alpha-ref-typing-full} defines the typing rules of
  \LAlpha{}.
\item Figure~\ref{fig:runtime-lambda-alpha-ref-types} defines the typing rules
  for the runtime language of \LAlpha{}.
\item Figure~\ref{fig:lambda-alpha-ref-reduction-full} defines the reduction
  semantics of \LAlpha{}.
\item Figure~\ref{fig:lctx-wf} defines the well-formedness for code contexts and
  types of \LCtx{}.
\item Figure~\ref{fig:lctx-typing-full} defines the typing rules of \LCtx{}.
\item Figure~\ref{fig:lctx-semantics-full} defines the reduction semantics of
  \LCtx{}.
\item Figure~\ref{fig:compile-types-full} defines the compilation of classifiers
  and types.
\item Figure~\ref{fig:compile-subtyping-full} defines the compilation of
  subtyping.
\item Figure~\ref{fig:compile-terms-full} defines the compilation of code terms
  and meta terms.
\item Figure~\ref{fig:compile-ctx} maps a compilation context to a typing
  context of \LAlpha{} and to a meta typing context of \LCtx{}.
\item Definition~\ref{def:wf-ctx} defines well-formed compilation contexts.
\item Figure~\ref{fig:reify} defines the reification of the code values of
  \LAlpha{} into STLC terms.
\item Figure~\ref{fig:sim-code} defines the simulation relation on
  configurations, heaps, and code terms.
\item Figure~\ref{fig:sim-meta} defines the simulation relation on meta terms.
\item Figure~\ref{fig:sim-coerce} defines the coerced value relation that the
  simulation relation uses.
\end{enumerate}

\clearpage

\begin{table}[H]
\centering
\footnotesize
\setlength{\tabcolsep}{2pt}
\begin{tabular}{@{}>{\raggedright\arraybackslash}p{0.27\textwidth}
                >{\raggedright\arraybackslash}p{0.35\textwidth}
                >{\raggedright\arraybackslash}p{0.35\textwidth}@{}}
\toprule
\textbf{\normalsize Results} & \textbf{\normalsize Agda identifiers} & \textbf{\normalsize Agda files} \\
\midrule
\multicolumn{3}{@{}l}{\textbf{\normalsize Section~\ref{sec:type-preservation}: Compilation preserves types}}\\
Well-formed compilation context (Definition~\ref{def:wf-ctx}) &
  \texttt{CompileCtxWF} &
  \path{CompileCtx.agda} \\
Compiled contexts and types are well formed (Lemma~\ref{lem:ec-unique}) &
  \texttt{ec-compiles},
  \texttt{ec-unique},
  \texttt{ec-wf},
  \texttt{type-compiles},
  \texttt{type-unique},
  \texttt{type-wf} &
  \path{Compile.agda},
  \path{CompilePres.agda} \\
Subtyping compiles to well-typed terms (Lemma~\ref{lem:sub-pres}) &
  \texttt{compile-ec-subtyping-}\allowbreak\texttt{preserves-types},
  \texttt{compile-subtyping-}\allowbreak\texttt{preserves-types} &
  \path{CompilePres.agda} \\
Compilation preserves types under a compilation context
  (Lemma~\ref{lem:compile-pres}) &
  \texttt{compile-code},
  \texttt{compile-meta} &
  \path{CompilePres.agda} \\
Compilation preserves types (Theorem~\ref{thm:compile-pres-types}) &
  \texttt{compile-preserves-types} &
  \path{CompilePres.agda} \\
\addlinespace
\multicolumn{3}{@{}l}{\textbf{\normalsize Section~\ref{sec:semantic-correctness}: Semantic correctness}}\\
Simulation relation (Figures~\ref{fig:sim-code}--\ref{fig:sim-coerce}) &
  \texttt{\_;\_;\_}$\vdash$\textsuperscript{\texttt{o}}\texttt{\_:\_}$\sim$\texttt{\_},
  \texttt{\_;\_}$\vdash$\textsuperscript{\texttt{m}}\texttt{\_:\_}$\sim$\texttt{\_},
  \texttt{\_}$\vdash$\texttt{\_<:\_}$\leadsto$\texttt{\_}$\Rightarrow$\texttt{\_} &
  \path{Simulation/Relation.agda},\newline
  \path{Simulation/Subtyping.agda} \\
Compilation embeds into simulation (Lemma~\ref{lem:sim-embed}) &
  \texttt{embed-code-compilation},
  \texttt{embed-meta-compilation} &
  \path{Simulation/Relation.agda} \\
Forward simulation (Lemma~\ref{lem:forward-sim}) &
  \texttt{forward-simulation}\textsuperscript{\texttt{o}},
  \texttt{forward-simulation},
  \texttt{forward-simulation*} &
  \path{ForwardSimulation.agda} \\
Backward simulation (Lemma~\ref{lem:backward-sim}) &
  \texttt{backward-simulation}\textsuperscript{\texttt{o}},
  \texttt{backward-simulation},
  \texttt{backward-simulation*} &
  \path{BackwardSimulation.agda} \\
Semantic correctness (Theorem~\ref{thm:semantic-correctness}) &
  \texttt{semantic-preservation},
  \texttt{semantic-reflection} &
  \path{CompilerCorrectness.agda} \\
\bottomrule
\end{tabular}
\caption{Mapping from metatheoretical results to their Agda mechanization. File
  paths are relative to the root of the Agda development (the
  \texttt{Theories/} directory).}
\label{tab:paper-agda-result-map}
\end{table}

\clearpage

\section{Auxiliary Definitions for \LAlpha}

\begin{figure}[H]
\footnotesize
\hfill\Fbox{$\Gamma \vdash e$}
\begin{gather*}
  \inference{}
            {\Gamma \vdash \ECeps}
  \quad
  \inference{\ECalpha \in \Gamma}
            {\Gamma \vdash \ECalpha}
\end{gather*}

\hfill\Fbox{$\Gamma \vdash A$}
\begin{gather*}
  \inference{}
            {\Gamma \vdash \iota}
  \quad
  \inference{\Gamma \vdash A
             & \Gamma \vdash B}
            {\Gamma \vdash A \to B}
  \quad
  \inference{\Gamma \vdash A}
            {\Gamma \vdash \REFTY{A}}
  \\
  \inference{\Gamma \vdash e}
            {\Gamma \vdash \CodeQuote{T}e}
  \quad
  \inference{\Gamma,\ECalpha \vdash A & \ECalpha \notin \Gamma }
            {\Gamma \vdash \forall \ECalpha.\,A}
  \quad
  \inference{\Gamma \vdash e_1
             & \Gamma \vdash e_2
             & \Gamma \vdash A}
            {\Gamma \vdash e_1{<:}e_2{\Rightarrow}A}
\end{gather*}
\caption{Well-formedness for ECs and meta types of \LAlpha{}}
\label{fig:lambda-alpha-ref-well-formedness}
\end{figure}

\begin{figure}[H]
\footnotesize
\hfill\Fbox{$\Theta \vdash A <: B$}
\begin{gather*}
  \inference{}
            {\Theta \vdash \iota <: \iota}
  \RuleTag{${<:}$base}
  \qquad
  \inference{\Theta \vdash A_2 <: A_1
             & \Theta \vdash B_1 <: B_2}
            {\Theta \vdash A_1 \to B_1 <: A_2 \to B_2}
  \RuleTag{${<:}$fun}
  \\[0.5ex]
  \inference{\Theta \vdash A <: B
             & \Theta \vdash B <: A}
            {\Theta \vdash \REFTY{A} <: \REFTY{B}}
  \RuleTag{${<:}$ref}
  \qquad
  \inference{\Theta \vdash e_1 <: e_2}
            {\Theta \vdash \CodeQuote{T}e_1 <: \CodeQuote{T}e_2}
  \RuleTag{${<:}$code}
  \\[0.5ex]
  \inference{\Theta \vdash A <: B}
            {\Theta \vdash \forall \ECalpha.\,A <: \forall \ECalpha.\,B}
  \RuleTag{${<:}$all}
  \qquad
  \inference{\Theta,e_1{<:}e_2 \vdash A <: B}
            {\Theta \vdash e_1{<:}e_2{\Rightarrow}A
             <: e_1{<:}e_2{\Rightarrow}B}
  \RuleTag{${<:}$sub}
\end{gather*}
\caption{Subtyping rules of \LAlpha{}.}
\label{fig:lambda-alpha-ref-subtyping-full}
\end{figure}

\begin{figure}[H]
\footnotesize
\hfill\Fbox{$\Gamma;e \vdash M^\circ : T$}
\begin{gather*}
  \inference{k : \iota}
            {\Gamma;e \vdash k : \iota}
  \RuleTag{${\vdash}$code-const}
  \qquad
  \inference{(x:T)^{\ECalpha} \in \Gamma}
            {\Gamma;\ECalpha \vdash x : T}
  \RuleTag{${\vdash}$code-var}
  \\[0.5ex]
  \inference{\Gamma,\ECalpha,e{<:}\ECalpha,(x:S)^{\ECalpha};\ECalpha
             \vdash N^\circ : T
            \\ \ECalpha \notin \Gamma}
            {\Gamma;e \vdash \lambda (x{:}S)^{\ECalpha}.\,N^\circ : S \to T}
  \RuleTag{${\vdash}$code-abs}
  \qquad
  \inference{\Gamma;e \vdash L^\circ : S \to T
             \\
             \Gamma;e \vdash M^\circ : S}
            {\Gamma;e \vdash L^\circ\;M^\circ : T}
  \RuleTag{${\vdash}$code-app}
  \\[0.5ex]
  \inference{\Gamma \vdash M : \CodeQuote{T}e}
            {\Gamma;e \vdash \SPLICE{M} : T}
  \RuleTag{${\vdash}$splice}
  \qquad
  \inference{\Gamma;e_1 \vdash M^\circ : T
            & \mathsf{Sub}(\Gamma) \vdash e_1 <: e_2}
            {\Gamma;e_2 \vdash M^\circ : T}
  \RuleTag{${\vdash}$code-subenv}
\end{gather*}

\hfill\Fbox{$\Gamma \vdash M : A$}
\begin{gather*}
  \inference{k : \iota}
            {\Gamma \vdash k : \iota}
  \RuleTag{${\vdash}$meta-const}
  \quad
  \inference{x:A \in \Gamma}
            {\Gamma \vdash x : A}
  \RuleTag{${\vdash}$meta-var}
  \quad
  \inference{\Gamma \vdash A
             \\
             \Gamma,x:A \vdash M : B}
            {\Gamma \vdash \lambda x{:}A.\,M : A \to B}
  \RuleTag{${\vdash}$meta-abs}
  \\[0.5ex]
  \inference{\Gamma \vdash L : A \to B
             & \Gamma \vdash M : A}
            {\Gamma \vdash L\;M : B}
  \RuleTag{${\vdash}$meta-app}
  \quad
  \inference{\Gamma,\ECalpha \vdash M : A &
             \ECalpha \notin \Gamma}
            {\Gamma \vdash \Lambda \ECalpha.\,M : \forall \ECalpha.\,A}
  \RuleTag{${\vdash}$ec-abs}
  \\[0.5ex]
  \inference{\Gamma \vdash M : \forall \ECalpha.\,A
             & \Gamma \vdash e}
            {\Gamma \vdash M[e] : A[\ECalpha := e]}
  \RuleTag{${\vdash}$ec-app}
  \quad
  \inference{\Gamma \vdash e_1
             &
             \Gamma \vdash e_2
             \\
             \Gamma,e_1{<:}e_2 \vdash M : A}
            {\Gamma \vdash e_1{<:}e_2{\Rightarrow}M :
             e_1{<:}e_2{\Rightarrow}A}
  \RuleTag{${\vdash}$sub-intro}
  \\[0.5ex]
  \inference{\Gamma \vdash M : e_1{<:}e_2{\Rightarrow}A
             \\ \mathsf{Sub}(\Gamma) \vdash e_1 <: e_2}
            {\Gamma \vdash M\bullet : A}
  \RuleTag{${\vdash}$sub-elim}
  \quad
  \inference{\Gamma \vdash e
             \\
             \Gamma;e \vdash M^\circ : T}
            {\Gamma \vdash \CodeQuote{M^\circ}e : \CodeQuote{T}e}
  \RuleTag{${\vdash}$quote}
  \\[0.5ex]
  \inference{\Gamma \vdash M : A}
            {\Gamma \vdash \REF{M} : \REFTY{A}}
  \RuleTag{${\vdash}$ref}
  \quad
  \inference{\Gamma \vdash L : \REFTY{A}
             & \Gamma \vdash M : A}
            {\Gamma \vdash L := M : \UNIT}
  \RuleTag{${\vdash}$assign}
  \\[0.5ex]
  \inference{\Gamma \vdash M : \REFTY{A}}
            {\Gamma \vdash \mathop{!}M : A}
  \RuleTag{${\vdash}$deref}
  \quad
  \inference{\Gamma \vdash M : A
             & \Gamma \vdash B
             & \mathsf{Sub}(\Gamma) \vdash A <: B}
            {\Gamma \vdash M : B}
  \RuleTag{${\vdash}$meta-sub}
\end{gather*}
\caption{Typing rules of \LAlpha{}.}
\label{fig:lambda-alpha-ref-typing-full}
\end{figure}

\begin{figure}[H]
\footnotesize
\setlength{\abovedisplayskip}{2pt}\setlength{\belowdisplayskip}{2pt}%
\hfill\Fbox{$\Gamma;\Sigma;e \vdash M^\circ : T$}
\begin{gather*}
  \dots
  \qquad
  \inference{\RuleHighlight{\Gamma \vdash \ECalpha}
             & \RuleHighlight{(x:S)^{\ECalpha} \in \Gamma}
             & \RuleHighlight{e <: \ECalpha \in \Gamma}
             & \Gamma;\Sigma;\ECalpha \vdash N^\circ : T}
            {\Gamma;\Sigma;e \vdash \overline{\lambda} (x{:}S)^{\ECalpha}.\,N^\circ : S \to T}
  \RuleTag{${\vdash}$code-abs-val}
\end{gather*}
\hfill\Fbox{$\Gamma;\Sigma \vdash M : A$}
\begin{gather*}
  \dots
  \qquad
  \inference{a:A \in \Sigma}
            {\Gamma;\Sigma \vdash a : \REFTY{A}}
  \RuleTag{${\vdash}$addr}
\end{gather*}
\caption{Typing rules for the runtime language of \LAlpha{}}
\label{fig:runtime-lambda-alpha-ref-types}
\end{figure}

\clearpage

\begin{figure}[H]
\footnotesize
\[
\begin{array}{lrcl}
  \text{code frames} & F^\circ & ::= &
     \Box\;M^\circ \mid V^\circ\;\Box\\
  \text{meta frames} & F & ::= &
     \Box\;M \mid V\;\Box \mid \REF{\Box}
     \mid \Box := M \mid V := \Box \mid \mathop{!}\Box
     \mid \Box[e] \mid \Box\bullet
\end{array}
\]

\hfill\Fbox{$
  e \vdash \langle\Gamma;M^\circ;\mu\rangle
  \longrightarrow^\circ
  \langle\Gamma';N^\circ;\mu'\rangle
$}
\begin{gather*}
  \inference{
    e \vdash \langle\Gamma;M^\circ;\mu\rangle 
    \longrightarrow^\circ 
    \langle\Gamma';N^\circ;\mu'\rangle}
    {
      e \vdash \langle\Gamma;F^\circ[M^\circ];\mu\rangle
      \longrightarrow^\circ
      \langle\Gamma';F^\circ[N^\circ];\mu'\rangle}
  \RuleTag{$\xi$-code}
  \\
  \inference{\langle\Gamma;M;\mu\rangle
              \longrightarrow^m
              \langle\Gamma';N;\mu'\rangle}
            {e \vdash \langle\Gamma;\SPLICE{M};\mu\rangle
              \longrightarrow^\circ
              \langle\Gamma';\SPLICE{N};\mu'\rangle}
  \RuleTag{$\xi$-splice}
  \\[2ex]
  \inference{}
  {e \vdash \langle\Gamma;\SPLICE{\CodeQuote{V^\circ}e'};\mu\rangle
    \longrightarrow^\circ
    \langle\Gamma;V^\circ;\mu\rangle}
  \RuleTag{$\beta$-splice}
  \\[2ex]
  \inference{\ECbeta,y\notin\Gamma}
  {\begin{array}{c}
             e\vdash\langle \Gamma;\lambda(x{:}T)^{\ECalpha}.\,N^\circ;\mu\rangle
             \longrightarrow^\circ
             \langle (\Gamma,\ECbeta,(e{<:}\ECbeta),
               (y{:}T)^{\ECbeta});
               \overline{\lambda}(y{:}T)^{\ECbeta}.\,
               N^\circ[\ECalpha:=\ECbeta][x:=y];\mu\rangle
             \end{array}}
  \RuleTag{code-$\lambda$}
  \\[2ex]
  \inference{\ECalpha\vdash\langle \Gamma; 
               N^\circ;\mu\rangle 
               \longrightarrow^\circ 
               \langle \Gamma';N^{\circ\prime};\mu'\rangle
             }
            {
             e\vdash\langle \Gamma;
             \overline{\lambda}
             (x{:}T)^{\ECalpha}.
             N^\circ;\mu\rangle
             \longrightarrow^\circ 
             \langle \Gamma'; 
             \overline{\lambda} 
             (x{:}T)^{\ECalpha}. 
             N^{\circ\prime};\mu'\rangle
             }
  \RuleTag{$\xi$-code-$\bar{\lambda}$}
\end{gather*}

\hfill\Fbox{$
  \langle\Gamma;M;\mu\rangle
  \longrightarrow^m
  \langle\Gamma';N;\mu'\rangle
$}
\begin{gather*}
  \inference{\langle\Gamma;M;\mu\rangle
             \longrightarrow^m
             \langle\Gamma';N;\mu'\rangle}
            {\langle\Gamma;F[M];\mu\rangle
             \longrightarrow^m
             \langle\Gamma';F[N];\mu'\rangle}
  \RuleTag{$\xi$-meta}
  \\[1ex]
  \inference{e \vdash \langle\Gamma;M^\circ;\mu\rangle
             \longrightarrow^\circ
             \langle\Gamma';N^\circ;\mu'\rangle}
            {\langle\Gamma;\CodeQuote{M^\circ}e;\mu\rangle
             \longrightarrow^m
             \langle\Gamma';\CodeQuote{N^\circ}e;\mu'\rangle}
  \RuleTag{$\xi$-quote}
  \\[1.5ex]
  \inference{}
            {\langle\Gamma;(\lambda x{:}A.\,M)\;V;\mu\rangle
             \longrightarrow^m
             \langle\Gamma;M[x:=V];\mu\rangle}
  \RuleTag{$\beta$}
  \\[1ex]
  \inference{}
            {\langle\Gamma;(\Lambda \ECalpha.\,M)[e];\mu\rangle
             \longrightarrow^m
             \langle\Gamma;M[\ECalpha:=e];\mu\rangle}
  \RuleTag{$\beta$-ec}
  \\[1ex]
  \inference{}
            {\langle\Gamma;(e_1{<:}e_2{\Rightarrow}M)\bullet;\mu\rangle
             \longrightarrow^m
             \langle\Gamma;M;\mu\rangle}
  \RuleTag{$\beta$-sub}
  \\[1ex]
  \inference{a \notin \mathsf{dom}(\mu)}
            {\langle\Gamma;\REF{V};\mu\rangle
             \longrightarrow^m
             \langle\Gamma;a;\mu[a \mapsto V]\rangle}
  \RuleTag{ref}
  \\[1ex]
  \inference{\mu(a)=V}
            {\langle\Gamma;\mathop{!}a;\mu\rangle
             \longrightarrow^m
             \langle\Gamma;V;\mu\rangle}
  \RuleTag{deref}
  \\[1ex]
  \inference{}
            {
             \langle\Gamma; 
             a := V;\mu\rangle
             \longrightarrow^m 
             \langle\Gamma; 
             ();\mu[a \mapsto V]\rangle
             }
  \RuleTag{assign}
\end{gather*}
\caption{Operational semantics of \LAlpha{}.}
\label{fig:lambda-alpha-ref-reduction-full}
\end{figure}

\clearpage

\section{Auxiliary Definitions for \LCtx}

\begin{figure}[H]
\footnotesize
\hfill\Fbox{$\Gamma \vdash \Gamma^\circ \;\mathsf{ctx}$}
\begin{gather*}
  \inference{}{\Gamma \vdash \emptyset \;\mathsf{ctx}}
  \quad
  \inference{\alpha \in \Gamma}{\Gamma \vdash \alpha \;\mathsf{ctx}}
  \quad
  \inference{}{\Gamma \vdash x{:}T \;\mathsf{ctx}}
  \quad
  \inference{\Gamma \vdash \Gamma^\circ \;\mathsf{ctx} & \Gamma \vdash \Delta^\circ \;\mathsf{ctx}}
            {\Gamma \vdash \Gamma^\circ, \Delta^\circ \;\mathsf{ctx}}
\end{gather*}
\hfill\Fbox{$\Gamma \vdash A \;\mathsf{ty}$}
\begin{gather*}
  \inference{}{\Gamma \vdash \iota \;\mathsf{ty}}
  \quad
  \inference{\Gamma \vdash A \;\mathsf{ty} & \Gamma \vdash B \;\mathsf{ty}}
            {\Gamma \vdash A \to B \;\mathsf{ty}}
  \quad
  \inference{\Gamma \vdash A \;\mathsf{ty}}{\Gamma \vdash \REFTY{A} \;\mathsf{ty}}
  \quad
  \inference{\Gamma \vdash \Gamma^\circ \;\mathsf{ctx}}
            {\Gamma \vdash \CQuote{\Gamma^\circ}{T} \;\mathsf{ty}}
  \\[1ex]
  \inference{\Gamma, \alpha \vdash A \;\mathsf{ty}}
            {\Gamma \vdash \forall \alpha.\,A \;\mathsf{ty}}
  \qquad
  \inference{\Gamma \vdash \Gamma^\circ \;\mathsf{ctx} & \Gamma \vdash \Delta^\circ \;\mathsf{ctx}}
            {\Gamma \vdash \SubstTy{\Gamma^\circ}{\Delta^\circ} \;\mathsf{ty}}
\end{gather*}
\caption{Well-formedness of code contexts and meta types in \LCtx.}
\label{fig:lctx-wf}
\end{figure}

Figure~\ref{fig:lctx-wf} defines the well-formedness judgments for code contexts
and meta types of \LCtx.
A code context is well formed when every context variable in it is well scoped
in $\Gamma$.
Three rules for meta types concern context variables and code contexts. The rule
for $\forall \alpha.\,A$ binds the context variable $\alpha$ in the body $A$.
The rule for code types $\CQuote{\Gamma^\circ}{T}$ requires $\Gamma^\circ$ to be
well formed. The rule for substitution types
$\SubstTy{\Gamma^\circ}{\Delta^\circ}$ requires both the domain $\Gamma^\circ$
and the codomain $\Delta^\circ$ to be well formed.

\clearpage

\begin{figure}[H]
\footnotesize
\hfill\Fbox{$x{:}T \in \Gamma^\circ$}
\begin{gather*}
  \inference{}{x{:}T \in x{:}T}
  \qquad
  \inference{x{:}T \in \Gamma^\circ}{x{:}T \in (\Gamma^\circ, \Delta^\circ)}
  \qquad
  \inference{x{:}T \in \Delta^\circ}{x{:}T \in (\Gamma^\circ, \Delta^\circ)}
\end{gather*}
\hfill\Fbox{$\Gamma; \Sigma; \Gamma^\circ \vdash M^\circ : T$}
\begin{gather*}
  \inference{k : \iota}
            {\Gamma; \Sigma; \Gamma^\circ \vdash k : \iota}
  \RuleTag{${\vdash}$code-const}
  \quad
  \inference{x{:}T \in \Gamma^\circ}
            {\Gamma; \Sigma; \Gamma^\circ \vdash x : T}
  \RuleTagHighlight{${\vdash}$code-var}
  \\[1ex]
  \inference{\Gamma; \Sigma; (\Gamma^\circ, x{:}S) \vdash N^\circ : T}
            {\Gamma; \Sigma; \Gamma^\circ \vdash \lambda (x{:}S).\,N^\circ : S \to T}
  \RuleTag{${\vdash}$code-abs}
  \quad
  \inference{\Gamma; \Sigma; \Gamma^\circ \vdash L^\circ : S \to T \\
             \Gamma; \Sigma; \Gamma^\circ \vdash M^\circ : S}
            {\Gamma; \Sigma; \Gamma^\circ \vdash L^\circ\;M^\circ : T}
  \RuleTag{${\vdash}$code-app}
  \\[1ex]
  \inference{\Gamma; \Sigma \vdash M : \CQuote{\Gamma^\circ_1}{T}
             & \Gamma; \Sigma \vdash N : \SubstTy{\Gamma^\circ_1}{\Gamma^\circ_2}}
            {\Gamma; \Sigma; \Gamma^\circ_2 \vdash \SPLICE{M}\,N : T}
  \RuleTagHighlight{${\vdash}$splice}
\end{gather*}
\hfill\Fbox{$\Gamma; \Sigma \vdash M : A$}
\begin{gather*}
  \inference{k : \iota}
            {\Gamma; \Sigma \vdash k : \iota}
  \RuleTag{${\vdash}$const}
  \quad
  \inference{x{:}A \in \Gamma}
            {\Gamma; \Sigma \vdash x : A}
  \RuleTag{${\vdash}$var}
  \quad
  \inference{a{:}A \in \Sigma}
            {\Gamma; \Sigma \vdash a : \REFTY{A}}
  \RuleTag{${\vdash}$addr}
  \\[1ex]
  \inference{\Gamma \vdash A \;\mathsf{ty}
             & \Gamma, x{:}A; \Sigma \vdash N : B}
            {\Gamma; \Sigma \vdash \lambda (x{:}A).\,N : A \to B}
  \RuleTag{${\vdash}$abs}
  \;
  \inference{\Gamma; \Sigma \vdash L : A \to B
             & \Gamma; \Sigma \vdash M : A}
            {\Gamma; \Sigma \vdash L\;M : B}
  \RuleTag{${\vdash}$app}
  \\[1ex]
  \inference{\Gamma \vdash \Gamma^\circ \;\mathsf{ctx}
             & \Gamma; \Sigma; \Gamma^\circ \vdash M^\circ : T}
            {\Gamma; \Sigma \vdash \CQuote{\Gamma^\circ}{M^\circ} : \CQuote{\Gamma^\circ}{T}}
  \RuleTagHighlight{${\vdash}$quote}
  \\[1ex]
  \inference{\Gamma, \alpha; \Sigma \vdash N : A & \alpha \notin \Gamma}
            {\Gamma; \Sigma \vdash \Lambda \alpha.\,N : \forall \alpha.\,A}
  \RuleTagHighlight{${\vdash}$ctx-abs}
  \;
  \inference{\Gamma; \Sigma \vdash M : \forall \alpha.\,A
             & \Gamma \vdash \Gamma^\circ \;\mathsf{ctx}}
            {\Gamma; \Sigma \vdash M[\Gamma^\circ] : A[\alpha := \Gamma^\circ]}
  \RuleTagHighlight{${\vdash}$ctx-app}
  \\[1ex]
  \inference{\Gamma; \Sigma \vdash M : A}
            {\Gamma; \Sigma \vdash \REF{M} : \REFTY{A}}
  \RuleTag{${\vdash}$ref}
  \qquad
  \inference{\Gamma; \Sigma \vdash M : \REFTY{A}}
            {\Gamma; \Sigma \vdash \mathop{!}M : A}
  \RuleTag{${\vdash}$deref}
  \\[1ex]
  \inference{\Gamma; \Sigma \vdash L : \REFTY{A} \\
             \Gamma; \Sigma \vdash M : A}
            {\Gamma; \Sigma \vdash L := M : \Unit}
  \RuleTag{${\vdash}$assign}
  \qquad
  \inference{\Gamma; \Sigma \vdash L : \REFTY{A} \\
             \Gamma; \Sigma \vdash M : A \to B   \\
             \Gamma; \Sigma \vdash N : B \to A}
            {\Gamma; \Sigma \vdash \Proxy{L}{M}{N} : \REFTY{B}}
  \RuleTagHighlight{${\vdash}$proxy}
  \\[1ex]
  \inference{\Gamma \vdash \Gamma^\circ \;\mathsf{ctx}}
            {\Gamma; \Sigma \vdash \IdSub{\Gamma^\circ} : \SubstTy{\Gamma^\circ}{\Gamma^\circ}}
  \RuleTagHighlight[GreenYellow!50]{${\vdash}$id}
  \;
  \inference{\Gamma \vdash \Gamma^\circ \;\mathsf{ctx}}
            {\Gamma; \Sigma \vdash \WeakenSub{\Gamma^\circ}{\IdSub{\emptyset}} : \SubstTy{\emptyset}{\Gamma^\circ}}
  \RuleTagHighlight[GreenYellow!50]{${\vdash}$weaken-empty}
  \\[1ex]
  \inference{\Gamma; \Sigma \vdash M : \SubstTy{\Gamma^\circ_1}{\Gamma^\circ_2} &
             \Gamma \vdash \Gamma^\circ_3 \;\mathsf{ctx}}
            {\Gamma; \Sigma \vdash \WeakenSub{\Gamma^\circ_3}{M} : \SubstTy{\Gamma^\circ_1}{(\Gamma^\circ_2, \Gamma^\circ_3)}}
  \RuleTagHighlight[GreenYellow!50]{${\vdash}$weaken}
  \\[1ex]
  \inference{\Gamma; \Sigma \vdash M : \SubstTy{\Gamma^\circ}{\Delta^\circ} &
             \Gamma; \Sigma; \Delta^\circ \vdash N^\circ : T}
            {\Gamma; \Sigma \vdash (M, x{=}N^\circ) : \SubstTy{(\Gamma^\circ, x{:}T)}{\Delta^\circ}}
  \RuleTagHighlight[GreenYellow!50]{${\vdash}$ext}
  \\[1ex]
  \inference{\Gamma; \Sigma \vdash M : \SubstTy{\Gamma^\circ_1}{\Gamma^\circ_2} &
             \Gamma; \Sigma \vdash N : \SubstTy{\Gamma^\circ_2}{\Gamma^\circ_3}}
            {\Gamma; \Sigma \vdash M \circ N : \SubstTy{\Gamma^\circ_1}{\Gamma^\circ_3}}
  \RuleTagHighlight[GreenYellow!50]{${\vdash}$compose}
\end{gather*}
\caption{Typing rules of \LCtx.}
\label{fig:lctx-typing-full}
\end{figure}

\clearpage

\begin{figure}[H]
\footnotesize
\setlength{\abovedisplayskip}{2pt}\setlength{\belowdisplayskip}{2pt}%
\[
\begin{array}{lrcl}
\text{code values} & V^\circ, W^\circ & ::= &
  k \mid x \mid \lambda (x{:}T).\,V^\circ \mid V^\circ\;W^\circ \\
\text{meta values} & V, W & ::= &
  k \mid a \mid \lambda (x{:}A).\,N \mid \Lambda \alpha.\,N
  \mid \CQuote{\Gamma^\circ}{V^\circ} \mid \Proxy{V}{V}{V} \\
  & & \mid & \IdSub{\Gamma^\circ} \mid (V, x{=}V^\circ)
  \mid \WeakenSub{\Gamma^\circ}{V} \mid V \circ W \\
\text{heaps} & \mu & ::= & \emptyset \mid \mu, a \mapsto V \\[1ex]
\text{code frames} & F^\circ & ::= &
  \Box\;M^\circ \mid V^\circ\;\Box \mid \lambda (x{:}T).\,\Box \\
\text{meta frames} & F & ::= &
  \Box\;M \mid V\;\Box \mid \Box[\Gamma^\circ]
  \mid \Box \circ M \mid V \circ \Box \\
  & & \mid & \WeakenSub{\Gamma^\circ}{\Box} \mid (\Box, x{=}V^\circ)
  \mid \REF{\Box} \mid \Box := M \mid V := \Box \\
  & & \mid & \mathop{!}\Box \mid \Proxy{\Box}{M}{N} \mid \Proxy{V}{\Box}{N} \mid \Proxy{V}{W}{\Box}
\end{array}
\]
\noindent
\begin{minipage}[t]{0.49\linewidth}
\hfill\Fbox{$\Sem{V}(x) = M^\circ$}
\begin{align*}
\Sem{\IdSub{\Gamma^\circ}}(x) &= x \\
\Sem{(V, y{=}V^\circ)}(x) &=
  \begin{cases}
    V^\circ   & \text{if } x = y \\
    \Sem{V}(x) & \text{if } x \neq y
  \end{cases} \\
\Sem{\WeakenSub{\Gamma^\circ}{V}}(x) &= \Sem{V}(x) \\
\Sem{V \circ W}(x) &= \Sem{W}(\Sem{V}(x))
\end{align*}
\end{minipage}%
\hfill
\begin{minipage}[t]{0.49\linewidth}
\hfill\Fbox{$\Sem{V}(M^\circ) = N^\circ$}
\begin{align*}
\Sem{V}(k) &= k \\
\Sem{V}(x) &= \Sem{V}(x) \\
\Sem{V}(\lambda (x{:}T).\,M^\circ) &= \lambda (x{:}T).\,\Sem{V}(M^\circ) \\
\Sem{V}(L^\circ\;M^\circ) &= (\Sem{V}(L^\circ))\;(\Sem{V}(M^\circ)) \\
\Sem{V}(\SPLICE{M}\,N) &= \SPLICE{M}\,N
\end{align*}
\end{minipage}\par
\hfill\Fbox{$M^\circ \mid \mu \longrightarrow^\circ N^\circ \mid \mu'$}
\begin{gather*}
  \inference{M^\circ \mid \mu \longrightarrow^\circ N^\circ \mid \mu'}
            {F^\circ[M^\circ] \mid \mu \longrightarrow^\circ F^\circ[N^\circ] \mid \mu'}
  \RuleTag{$\xi$-code}
  \quad
  \inference{M \mid \mu \longrightarrow^m M' \mid \mu'}
            {\SPLICE{M}\,N \mid \mu \longrightarrow^\circ \SPLICE{M'}\,N \mid \mu'}
  \RuleTagHighlight{$\xi_1$-splice}
  \\[1ex]
  \inference{N \mid \mu \longrightarrow^m N' \mid \mu'}
            {\SPLICE{V}\,N \mid \mu \longrightarrow^\circ \SPLICE{V}\,N' \mid \mu'}
  \RuleTagHighlight{$\xi_2$-splice}
  \\[1ex]
  \inference{}
            {\SPLICE{\CQuote{\Gamma^\circ}{V^\circ}}\,V \mid \mu
             \longrightarrow^\circ \Sem{V}(V^\circ) \mid \mu}
  \RuleTagHighlight[GreenYellow!50]{$\beta$-splice}
\end{gather*}
\hfill\Fbox{$M \mid \mu \longrightarrow^m N \mid \mu'$}
\begin{gather*}
  \inference{M \mid \mu \longrightarrow^m N \mid \mu'}
            {F[M] \mid \mu \longrightarrow^m F[N] \mid \mu'}
  \RuleTag{$\xi$-meta}
  \\[1ex]
  \inference{M^\circ \mid \mu \longrightarrow^\circ N^\circ \mid \mu'}
            {(M, x{=}M^\circ) \mid \mu \longrightarrow^m (M, x{=}N^\circ) \mid \mu'}
  \RuleTagHighlight{$\xi$-ext}
  \\[1ex]
  \inference{M^\circ \mid \mu \longrightarrow^\circ N^\circ \mid \mu'}
            {\CQuote{\Gamma^\circ}{M^\circ} \mid \mu \longrightarrow^m \CQuote{\Gamma^\circ}{N^\circ} \mid \mu'}
  \RuleTagHighlight{$\xi$-quote}
  \\[1ex]
  \inference{}
            {(\lambda (x{:}A).\,N)\;V \mid \mu \longrightarrow^m N[x := V] \mid \mu}
  \RuleTag{$\beta$}
  \\[1ex]
  \inference{}
            {(\Lambda \alpha.\,N)[\Gamma^\circ] \mid \mu \longrightarrow^m N[\alpha := \Gamma^\circ] \mid \mu}
  \RuleTagHighlight{$\beta$-ctx}
  \;
  \inference{a \notin \mathsf{dom}(\mu)}
            {\REF{V} \mid \mu \longrightarrow^m a \mid \mu, a \mapsto V}
  \RuleTag{ref}
  \\[1ex]
  \inference{}
            {a := V \mid \mu \longrightarrow^m \unit \mid \mu[a \mapsto V]}
  \RuleTag{assign}
  \quad
  \inference{\mu(a) = V}
            {\mathop{!}a \mid \mu \longrightarrow^m V \mid \mu}
  \RuleTag{deref}
  \\[1ex]
  \inference{}
            {\mathop{!}(\Proxy{V_1}{V_2}{V_3}) \mid \mu \longrightarrow^m V_2\;(\mathop{!}V_1) \mid \mu}
  \RuleTagHighlight{proxy-deref}
  \\[1ex]
  \inference{}
            {(\Proxy{V_1}{V_2}{V_3}) := W \mid \mu \longrightarrow^m V_1 := (V_3\;W) \mid \mu}
  \RuleTagHighlight{proxy-assign}
\end{gather*}
\caption{Operational semantics of \LCtx.}
\label{fig:lctx-semantics-full}
\end{figure}

Evaluation in \LCtx is call by value and left to right, as the frames
(non-recursive evaluation contexts) of Figure~\ref{fig:lctx-semantics-full}
specify. The congruence rules of each level are grouped into a single rule
that reduces under a frame ($\xi$\textit{-code}, $\xi$\textit{-meta}).

\clearpage

\section{Auxiliary Definitions for the Compilation From \LAlpha to \LCtx}

\begin{figure}[H]
\footnotesize
\[
\begin{array}{lrcl}
\text{compilation contexts} & \Xi & ::= & \emptyset \mid \Xi, \ECalpha
  \mid \Xi, \RuleHighlight[GreenYellow!50]{\ECalpha_{\mathsf{poly}}} \mid \Xi, x{:}A
  \mid \Xi, \RuleHighlight[GreenYellow!50]{s : e_1 \sub e_2} \\
  & & \mid & \Xi, (x{:}T)^{\ECalpha} \mid \Xi, e_1 \sub e_2
\end{array}
\]
\hfill\Fbox{$\Xi \vdash e \compile \Gamma^\circ$}
\begin{gather*}
  \inference{}{\Xi \vdash \ECeps \compile \emptyset}
  \RuleTag{${\compile}$empty}
  \qquad
  \inference{\ECalpha_{\mathsf{poly}} \in \Xi}
            {\Xi \vdash \ECalpha \compile \alpha}
  \RuleTagHighlight{${\compile}$poly-ec}
  \\[1ex]
  \inference{\ECalpha \in \Xi & e \sub \ECalpha \in \Xi & (x{:}T)^{\ECalpha} \in \Xi
             & \Xi \vdash e \compile \Gamma^\circ}
            {\Xi \vdash \ECalpha \compile (\Gamma^\circ, x{:}T)}
  \RuleTagHighlight{${\compile}$ec}
\end{gather*}
\hfill\Fbox{$\Xi \vdash A \compile A'$}
\begin{gather*}
  \inference{}{\Xi \vdash \iota \compile \iota}
  \RuleTag{${\compile}$base}
  \quad
  \inference{\Xi \vdash A \compile A' & \Xi \vdash B \compile B'}
            {\Xi \vdash (A \to B) \compile (A' \to B')}
  \RuleTag{${\compile}$fun}
  \\[1ex]
  \inference{\Xi \vdash A \compile A'}
            {\Xi \vdash \REFTY{A} \compile \REFTY{A'}}
  \RuleTag{${\compile}$ref}
  \quad
  \inference{\Xi \vdash e \compile \Gamma^\circ}
            {\Xi \vdash \Quote{T}{e} \compile \CQuote{\Gamma^\circ}{T}}
  \RuleTagHighlight{${\compile}$code}
  \\[1ex]
  \inference{\Xi, \ECalpha_{\mathsf{poly}} \vdash A \compile A' & \ECalpha \notin \Xi}
            {\Xi \vdash \forall \ECalpha.\,A \compile \forall \alpha.\,A'}
  \RuleTag{${\compile}$forall}
  \\[1ex]
  \inference{\Xi \vdash e_1 \sub e_2 \compile B
             & \Xi, s : e_1 \sub e_2 \vdash A \compile A' & s \notin \Xi}
            {\Xi \vdash (e_1 \sub e_2 \Rightarrow A) \compile (B \to A')}
  \RuleTagHighlight{${\compile}$constraint}
\end{gather*}
\caption{Compilation contexts, and compilation of classifiers and types.}
\label{fig:compile-types-full}
\end{figure}

\clearpage

\begin{figure}[H]
\footnotesize
\hfill\Fbox{$\Xi \vdash e_1 \sub e_2 \compile A$}
\begin{gather*}
  \inference{\Xi \vdash e_1 \compile \Gamma^\circ_1 & \Xi \vdash e_2 \compile \Gamma^\circ_2}
            {\Xi \vdash e_1 \sub e_2 \compile \SubstTy{\Gamma^\circ_1}{\Gamma^\circ_2}}
  \RuleTag{${\compile}$sub-ec}
\end{gather*}
\hfill\Fbox{$\Xi \vdash e_1 \sub e_2 \compile M$}
\begin{gather*}
  \inference{\Xi \vdash e \compile \Gamma^\circ}
            {\Xi \vdash \ECeps \sub e \compile \WeakenSub{\Gamma^\circ}{\IdSub{\emptyset}}}
  \RuleTag{${\compile}$sub-empty}
  \quad
  \inference{\Xi \vdash e \compile \Gamma^\circ}
            {\Xi \vdash e \sub e \compile \IdSub{\Gamma^\circ}}
  \RuleTag{${\compile}$sub-refl}
  \\[1ex]
  \inference{e \sub \ECalpha \in \Xi & (x{:}T)^{\ECalpha} \in \Xi \\ \Xi \vdash e \compile \Gamma^\circ}
            {\Xi \vdash e \sub \ECalpha \compile \WeakenSub{x{:}T}{\IdSub{\Gamma^\circ}}}
  \RuleTag{${\compile}$sub-ax}
  \quad
  \inference{(s : e_1 \sub e_2) \in \Xi}
            {\Xi \vdash e_1 \sub e_2 \compile s}
  \RuleTag{${\compile}$sub-var}
  \\[1ex]
  \inference{\Xi \vdash e_1 \sub e_3 \compile M & \Xi \vdash e_3 \sub e_2 \compile N}
            {\Xi \vdash e_1 \sub e_2 \compile M \circ N}
  \RuleTag{${\compile}$sub-trans}
\end{gather*}
\hfill\Fbox{$\Xi \vdash A \sub B \compile M$}
\begin{gather*}
  \inference{}
            {\Xi \vdash \iota \sub \iota \compile \lambda (x{:}\iota).\,x}
  \RuleTag{${\compile}$sub-base}
  \\[1ex]
  \inference{\Xi \vdash e_1 \compile \Gamma^\circ & \Xi \vdash e_2 \compile \Delta^\circ
             & \RuleHighlight[GreenYellow!50]{\Xi \vdash e_1 \sub e_2 \compile M}}
            {\Xi \vdash \Quote{T}{e_1} \sub \Quote{T}{e_2} \compile
             \lambda (c{:}\CQuote{\Gamma^\circ}{T}).\,\CQuote{\Delta^\circ}{\SPLICE{c}\,\RuleHighlight[GreenYellow!50]{M}}}
  \RuleTag{${\compile}$sub-code}
  \\[1ex]
  \inference{\Xi \vdash A_1 \compile A_1' & \Xi \vdash A_2 \compile A_2' & \Xi \vdash B_1 \compile B_1' \\
             \Xi \vdash B_1 \sub A_1 \compile L & \Xi \vdash A_2 \sub B_2 \compile M}
            {
             \Xi \vdash A_1 \to A_2 \sub B_1 \to B_2 \compile
             \lambda (f{:}A_1' \to A_2').\,\lambda (x{:}B_1').\,M\;(f\;(L\;x))
             }
  \RuleTag{${\compile}$sub-fun}
  \\[1ex]
  \inference{\Xi, \ECalpha_{\mathsf{poly}} \vdash A \compile A' & \Xi, \ECalpha_{\mathsf{poly}} \vdash A \sub B \compile M
             & \ECalpha \notin \Xi}
            {\Xi \vdash \forall \ECalpha.\,A \sub \forall \ECalpha.\,B \compile
             \lambda (f{:}\forall \alpha.\,A').\,\Lambda \alpha.\,M\;(f[\alpha])}
  \RuleTag{${\compile}$sub-forall}
  \\[1ex]
  \inference{\Xi \vdash A \compile A' & \Xi \vdash A \sub B \compile M & \Xi \vdash B \sub A \compile N}
            {\Xi \vdash \REFTY{A} \sub \REFTY{B} \compile \lambda (r{:}\REFTY{A'}).\,\Proxy{r}{M}{N}}
  \RuleTag{${\compile}$sub-ref}
  \\[1ex]
  \inference{\Xi \vdash e_1 \sub e_2 \compile C
             & \Xi, (s : e_1 \sub e_2) \vdash A \compile A' \\
             \Xi, (s : e_1 \sub e_2) \vdash A \sub B \compile M & s \notin \Xi}
            {\begin{array}[t]{@{}l@{}}
             \Xi \vdash (e_1 \sub e_2 \Rightarrow A) \sub (e_1 \sub e_2 \Rightarrow B) \compile {} \\
             \qquad\qquad \lambda (f{:}C \to A').\,\lambda (s{:}C).\,M\;(f\;s)
             \end{array}}
  \RuleTag{${\compile}$sub-constraint}
\end{gather*}
\caption{Compilation of subtyping between classifiers and between types.}
\label{fig:compile-subtyping-full}
\end{figure}

\clearpage

\begin{figure}[H]
\footnotesize
\setlength{\abovedisplayskip}{2pt}\setlength{\belowdisplayskip}{2pt}%
\hfill\Fbox{$\Xi; e \vdash M^\circ : T \compile M^{\circ\prime}$}
\begin{gather*}
  \inference{k : \iota}
            {\Xi; e \vdash k : \iota \compile k}
  \RuleTag{${\compile}$code-const}
  \quad
  \inference{(x{:}T)^{\ECalpha} \in \Xi}
            {\Xi; \ECalpha \vdash x : T \compile x}
  \RuleTag{${\compile}$code-var}
  \\[1ex]
  \inference{\Xi, \ECalpha, e \sub \ECalpha, (x{:}S)^{\ECalpha}; \ECalpha \vdash N^\circ : T \compile N^{\circ\prime}
             & \ECalpha \notin \Xi}
            {\Xi; e \vdash \lambda (x{:}S)^{\ECalpha}.\,N^\circ : S \to T \compile \lambda (x{:}S).\,N^{\circ\prime}}
  \RuleTag{${\compile}$code-abs}
  \\[1ex]
  \inference{\Xi; e \vdash L^\circ : S \to T \compile L^{\circ\prime} \\
             \Xi; e \vdash M^\circ : S \compile M^{\circ\prime}}
            {\Xi; e \vdash L^\circ\;M^\circ : T \compile L^{\circ\prime}\;M^{\circ\prime}}
  \RuleTag{${\compile}$code-app}
  \;
  \inference{\Xi \vdash M : \Quote{T}{e} \compile M' \\
             \Xi \vdash e \compile \Gamma^\circ}
            {\Xi; e \vdash \SPLICE{M} : T \compile \SPLICE{M'}\,\IdSub{\Gamma^\circ}}
  \RuleTag{${\compile}$splice}
  \\[1ex]
  \inference{\Xi; e_1 \vdash M^\circ : T \compile M^{\circ\prime}
             & \Xi \vdash e_1 \compile \Gamma^\circ
             & \RuleHighlight{\Xi \vdash e_1 \sub e_2 \compile M}}
            {\Xi; e_2 \vdash M^\circ : T \compile \SPLICE{\CQuote{\Gamma^\circ}{M^{\circ\prime}}}\,\RuleHighlight{M}}
  \RuleTagHighlight{${\compile}$subenv}
\end{gather*}
\hfill\Fbox{$\Xi \vdash M : A \compile M'$}
\begin{gather*}
  \inference{k : \iota}
            {\Xi \vdash k : \iota \compile k}
  \RuleTag{${\compile}$const}
  \quad
  \inference{x{:}A \in \Xi}
            {\Xi \vdash x : A \compile x}
  \RuleTag{${\compile}$var}
  \\[1ex]
  \inference{\Xi \vdash A \compile A' & \Xi, x{:}A \vdash N : B \compile N'}
            {\Xi \vdash \lambda (x{:}A).\,N : A \to B \compile \lambda (x{:}A').\,N'}
  \RuleTag{${\compile}$abs}
  \\[1ex]
  \inference{\Xi \vdash L : A \to B \compile L' \\ \Xi \vdash M : A \compile M'}
            {\Xi \vdash L\;M : B \compile L'\;M'}
  \RuleTag{${\compile}$app}
  \quad
  \inference{\Xi \vdash M : A \compile M'}
            {\Xi \vdash \REF{M} : \REFTY{A} \compile \REF{M'}}
  \RuleTag{${\compile}$ref}
  \\[1ex]
  \inference{\Xi \vdash L : \REFTY{A} \compile L' \\ \Xi \vdash M : A \compile M'}
            {\Xi \vdash L := M : \Unit \compile L' := M'}
  \RuleTag{${\compile}$assign}
  \quad
  \inference{\Xi \vdash M : \REFTY{A} \compile M'}
            {\Xi \vdash \mathop{!}M : A \compile \mathop{!}M'}
  \RuleTag{${\compile}$deref}
  \\[1ex]
  \inference{\Xi; e \vdash M^\circ : T \compile M^{\circ\prime} \\
             \Xi \vdash \RuleHighlight[GreenYellow!50]{e \compile \Gamma^\circ}}
            {\Xi \vdash \Quote{M^\circ}{{\RuleHighlight[GreenYellow!50]{e}}} : \Quote{T}{e} \compile
             \CQuote{\RuleHighlight[GreenYellow!50]{\Gamma^\circ}}{M^{\circ\prime}}}
  \RuleTagHighlight[GreenYellow!50]{${\compile}$quote}
  \\[1ex]
  \inference{\Xi, \ECalpha_{\mathsf{poly}} \vdash N : A \compile N' & \ECalpha \notin \Xi}
            {\Xi \vdash \Lambda \RuleHighlight[GreenYellow!50]{\ECalpha}.\,N : \forall \ECalpha.\,A \compile \Lambda \RuleHighlight[GreenYellow!50]{\alpha}.\,N'}
  \RuleTagHighlight[GreenYellow!50]{${\compile}$ec-abs}
  \\[1ex]
  \inference{\Xi \vdash M : \forall \ECalpha.\,A \compile M'
             & \Xi \vdash \RuleHighlight[GreenYellow!50]{e \compile \Gamma^\circ}}
            {\Xi \vdash M[\RuleHighlight[GreenYellow!50]{e}] : A[\ECalpha := e] \compile
             M'[\RuleHighlight[GreenYellow!50]{\Gamma^\circ}]}
  \RuleTagHighlight[GreenYellow!50]{${\compile}$ec-app}
  \\[1ex]
  \inference{\Xi \vdash e_1 \compile \Gamma^\circ_1 & \Xi \vdash e_2 \compile \Gamma^\circ_2 \\
             \Xi, \RuleHighlight{(s : e_1 \sub e_2)} \vdash N : A \compile N' & s \notin \Xi}
            {\Xi \vdash (e_1 \sub e_2 \Rightarrow N) : (e_1 \sub e_2 \Rightarrow A) \compile
             \lambda (\RuleHighlight{s{:}\SubstTy{\Gamma^\circ_1}{\Gamma^\circ_2}}).\,N'}
  \RuleTagHighlight{${\compile}$sub-intro}
  \\[1ex]
  \inference{\Xi \vdash M : (e_1 \sub e_2 \Rightarrow A) \compile M' \\ \RuleHighlight{\Xi \vdash e_1 \sub e_2 \compile N}}
            {\Xi \vdash M\bullet : A \compile M'\;N}
  \RuleTagHighlight{${\compile}$sub-elim}
  \\[1ex]
  \inference{\Xi \vdash M : A \compile M' \\ \RuleHighlight{\Xi \vdash A \sub B \compile L}}
            {\Xi \vdash M : B \compile L\;M'}
  \RuleTagHighlight{${\compile}$subsumption}
\end{gather*}
\caption{Compilation of code terms and meta terms.}
\label{fig:compile-terms-full}
\end{figure}

\clearpage

\section{Auxiliary Definitions for the Metatheoretical Results}

\begin{figure}[H]
\footnotesize
\hfill\Fbox{$\Erase{\Xi}$}
\[
\begin{array}{r@{\;=\;}l@{\qquad}r@{\;=\;}l}
\Erase{\emptyset} & \emptyset &
\Erase{\Xi, \ECalpha_{\mathsf{poly}}} & \Erase{\Xi}, \ECalpha \\
\Erase{\Xi, \ECalpha} & \Erase{\Xi}, \ECalpha &
\Erase{\Xi, s : e_1 \sub e_2} & \Erase{\Xi}, e_1 \sub e_2 \\
\Erase{\Xi, x{:}A} & \Erase{\Xi}, x{:}A &
\Erase{\Xi, (x{:}T)^{\ECalpha}} & \Erase{\Xi}, (x{:}T)^{\ECalpha} \\
\Erase{\Xi, e_1 \sub e_2} & \Erase{\Xi}, e_1 \sub e_2
\end{array}
\]
\hfill\Fbox{$\vdash \Xi \compile \Gamma$}
\begin{gather*}
  \inference{}{\vdash \emptyset \compile \emptyset}
  \RuleTag{${\compile}$ctx-empty}
  \quad
  \inference{\vdash \Xi \compile \Gamma}
            {\vdash \Xi, \ECalpha_{\mathsf{poly}} \compile \Gamma, \alpha}
  \RuleTag{${\compile}$ctx-poly}
  \\[1ex]
  \inference{\vdash \Xi \compile \Gamma & \Erase{\Xi} \vdash e
             & \ECalpha \notin \Xi}
            {\vdash \Xi, \ECalpha, e \sub \ECalpha, (x{:}T)^{\ECalpha} \compile \Gamma}
  \RuleTag{${\compile}$ctx-code}
  \\[1ex]
  \inference{\vdash \Xi \compile \Gamma \\ \Xi \vdash A \compile A'}
            {\vdash \Xi, x{:}A \compile \Gamma, x{:}A'}
  \RuleTag{${\compile}$ctx-var}
  \quad
  \inference{\vdash \Xi \compile \Gamma & \Xi \vdash e_1 \sub e_2 \compile A}
            {\vdash \Xi, (s : e_1 \sub e_2) \compile \Gamma, s{:}A}
  \RuleTag{${\compile}$ctx-sub-var}
\end{gather*}
\caption{Mappings from a compilation context to a typing context of the source
  language, \LAlpha, and a meta typing context of the target language, \LCtx.}
\label{fig:compile-ctx}
\end{figure}

Figure~\ref{fig:compile-ctx} defines the operator $\Erase{\Xi}$ and the
compilation $\vdash \Xi \compile \Gamma$ of a compilation context.

The operator $\Erase{\Xi}$ takes a compilation context and returns a \LAlpha
typing context that corresponds to it. The operator turns a polymorphic
classifier $\ECalpha_{\mathsf{poly}}$ into the classifier $\ECalpha$ and a named
constraint $s : e_1 \sub e_2$ into the constraint $e_1 \sub e_2$. It keeps every
other entry.

The compilation $\vdash \Xi \compile \Gamma$ turns a compilation context into a
\LCtx metalanguage typing context. The judgment turns a polymorphic classifier
into a context variable (${\compile}$\textit{ctx-poly}), gives a meta variable
its compiled type (${\compile}$\textit{ctx-var}), and gives the evidence
variable $s$ the substitution type that its subtype constraint compiles to
(${\compile}$\textit{ctx-sub-var}). A code abstraction adds three entries
together, namely its classifier, the subtype constraint to its parent $e$, and
its code variable. All three are dropped (${\compile}$\textit{ctx-code}).

\clearpage

The compilation lemmas of Section~\ref{sec:type-preservation} require
compilation contexts to be well formed.

\begin{definition}[Well-formed compilation context]
\label{def:wf-ctx}
\tightlist
A compilation context $\Xi$ is well formed, written $\vdash \Xi$, if the
following hold.
\begin{enumerate}
\setlength{\abovedisplayskip}{3pt}\setlength{\belowdisplayskip}{3pt}%
\setlength{\abovedisplayshortskip}{0pt}\setlength{\belowdisplayshortskip}{3pt}%
\item Every entry mentions only well-scoped classifiers and types:
  \[
  \begin{array}{l@{\quad}l}
  \forall\, x{:}A \in \Xi. & \Erase{\Xi} \vdash A \\
  \forall\, (s : e_1 \sub e_2) \in \Xi. &
    \Erase{\Xi} \vdash e_1 \text{ and } \Erase{\Xi} \vdash e_2 \\
  \forall\, (x{:}T)^{\ECalpha} \in \Xi. & \ECalpha \in \Xi \\
  \forall\, (e_1 \sub e_2) \in \Xi. &
    \Erase{\Xi} \vdash e_1 \text{ and } \exists\, \ECalpha \in \Xi.\; e_2 = \ECalpha
  \end{array}
  \]
\item Every classifier has at most one parent: \\
  if $e_1 \sub \ECalpha \in \Xi$ and $e_2 \sub \ECalpha \in \Xi$ then $e_1 = e_2$.
\item Every classifier points to at most one code variable: \\
  if $(x{:}T)^{\ECalpha} \in \Xi$ and $(y{:}S)^{\ECalpha} \in \Xi$
  then $x = y$ and $T = S$.
\end{enumerate}
\end{definition}

Reification turns a code value of \LAlpha into STLC.

\begin{figure}[H]
\footnotesize
\setlength{\abovedisplayskip}{2pt}\setlength{\belowdisplayskip}{2pt}%
\[
\begin{array}{lrcl}
  \text{STLC terms} & N & ::= & k \mid x \mid \lambda (x{:}T).\,N \mid N\;N \\
  \text{reification environments} & \rho & ::= & \emptyset \mid \rho, y \mapsto x
\end{array}
\]
\[
\begin{array}{rcl}
  \Reify{V^\circ} &=& \ReifyEnv{\emptyset}{V^\circ} \\
  \ReifyEnv{\rho}{k} &=& k \\
  \ReifyEnv{\rho}{y} &=& \rho(y) \\
  \ReifyEnv{\rho}{\overline{\lambda} (y{:}T)^{\ECbeta}.\,V^\circ}
    &=& \lambda (x{:}T).\,\ReifyEnv{\rho, y \mapsto x}{V^\circ} \\
  \ReifyEnv{\rho}{V^\circ_1\;V^\circ_2}
    &=& \ReifyEnv{\rho}{V^\circ_1}\;\ReifyEnv{\rho}{V^\circ_2}
\end{array}
\]
\caption{Reification of the code values of \LAlpha{} into STLC terms.}
\label{fig:reify}
\end{figure}

\clearpage

\begin{figure}[H]
\footnotesize
\setlength{\abovedisplayskip}{0pt}\setlength{\belowdisplayskip}{0pt}%
\[
\begin{array}{lrcl}
\text{simulation contexts} & \Psi & ::= &
  \emptyset \mid \Psi, \ECalpha \mid \Psi, (x{:}T)^{\ECalpha} \mid \Psi, e_1 \sub e_2 \\[0.5ex]
& \mathrm{dom}(\Psi) & = &
  \{\ECalpha \mid \ECalpha \in \Psi\} \cup \{x \mid (x{:}T)^{\ECalpha} \in \Psi\}
\end{array}
\medskip
\]
\hfill
\Fbox{$\Sigma; \Psi; e \vdash \langle\Gamma; M^\circ; \mu\rangle : T \simrel N^\circ \mid \mu'$}
\vspace{5pt}
\begin{gather*}
  \inference{\mathrm{dom}(\Gamma) = \mathrm{dom}(\Psi)
             & \Sigma; \Psi; e \vdash M^\circ : T \simrel N^\circ
             & \Sigma; \Psi \vdash \mu \simrel \mu'}
            {\Sigma; \Psi; e \vdash \langle\Gamma; M^\circ; \mu\rangle : T \simrel N^\circ \mid \mu'}
  \RuleTag{${\simrel}$code-config}
\end{gather*}
\hfill\Fbox{$\Sigma; \Psi \vdash \langle\Gamma; M; \mu\rangle : A \simrel N \mid \mu'$}
\vspace{5pt}
\begin{gather*}
  \inference{\mathrm{dom}(\Gamma) = \mathrm{dom}(\Psi)
             & \Sigma; \Psi \vdash M : A \simrel N
             & \Sigma; \Psi \vdash \mu \simrel \mu'}
            {\Sigma; \Psi \vdash \langle\Gamma; M; \mu\rangle : A \simrel N \mid \mu'}
  \RuleTag{${\simrel}$config}
\end{gather*}
\hfill\Fbox{$\Sigma; \Xi \vdash \mu \simrel \mu'$}
\begin{gather*}
  \inference{\mathrm{dom}(\Sigma) = \mathrm{dom}(\mu) = \mathrm{dom}(\mu') \\
             \forall\, (a{:}A) \in \Sigma.\; \Sigma; \Xi \vdash \mu(a) : A \simrel \mu'(a)}
            {\Sigma; \Xi \vdash \mu \simrel \mu'}
  \RuleTag{${\simrel}$heap}
\end{gather*}
\hfill\Fbox{$\Sigma; \Xi; e \vdash M^\circ : T \simrel N^\circ$}
\vspace{5pt}
\begin{gather*}
  \inference{k : \iota}
            {\Sigma; \Xi; e \vdash k : \iota \simrel k}
  \RuleTag{${\simrel}$code-const}
  \quad
  \inference{(x{:}T)^{\ECalpha} \in \Xi}
            {\Sigma; \Xi; \ECalpha \vdash x : T \simrel x}
  \RuleTag{${\simrel}$code-var}
  \\[1ex]
  \inference{\Sigma; \Xi, \ECalpha, e \sub \ECalpha, (x{:}S)^{\ECalpha}; \ECalpha
               \vdash N^\circ : T \simrel N^{\circ\prime}
             & \ECalpha \notin \Xi}
            {\Sigma; \Xi; e \vdash \lambda (x{:}S)^{\ECalpha}.\,N^\circ : S \to T
               \simrel \lambda (x{:}S).\,N^{\circ\prime}}
  \RuleTag{${\simrel}$code-abs}
  \\[1ex]
  \inference{\ECalpha \in \Xi & e \sub \ECalpha \in \Xi & (x{:}S)^{\ECalpha} \in \Xi \\
             \Sigma; \Xi; \ECalpha \vdash N^\circ : T \simrel N^{\circ\prime}}
            {\Sigma; \Xi; e \vdash \overline{\lambda} (x{:}S)^{\ECalpha}.\,N^\circ : S \to T
               \simrel \lambda (x{:}S).\,N^{\circ\prime}}
  \RuleTagHighlight{${\simrel}$code-abs-val}
  \\[1ex]
  \inference{\Sigma; \Xi; e \vdash L^\circ : S \to T \simrel L^{\circ\prime} &
             \Sigma; \Xi; e \vdash M^\circ : S \simrel M^{\circ\prime}}
            {\Sigma; \Xi; e \vdash L^\circ\;M^\circ : T \simrel L^{\circ\prime}\;M^{\circ\prime}}
  \RuleTag{${\simrel}$code-app}
  \\[1ex]
  \inference{\Sigma; \Xi \vdash M : \Quote{T}{e} \simrel M' &
             \Xi \vdash e \compile \Gamma^\circ}
            {\Sigma; \Xi; e \vdash (\SPLICE{M}) : T \simrel (\SPLICE{M'}\,\IdSub{\Gamma^\circ})}
  \RuleTag{${\simrel}$splice}
  \\[1ex]
  \inference{\Sigma; \Xi; e_1 \vdash M^\circ : T \simrel M^{\circ\prime}
             & \Xi \vdash e_1 \compile \Gamma^\circ
             & \Xi \vdash e_1 \sub e_2 \compile M}
            {\Sigma; \Xi; e_2 \vdash M^\circ : T \simrel (\SPLICE{\CQuote{\Gamma^\circ}{M^{\circ\prime}}}\,M)}
  \RuleTag{${\simrel}$subenv}
  \\[1ex]
  \inference{\Sigma; \Xi; e_1 \vdash V^\circ : T \simrel V^{\circ\prime} \\
             \Xi \vdash \Quote{T}{e_1} \sub \Quote{T}{e_2} \coerceval
               \CQuote{\Gamma^\circ}{V^{\circ\prime}} \Rightarrow \CQuote{\Delta^\circ}{W^{\circ\prime}}}
            {\Sigma; \Xi; e_2 \vdash V^\circ : T \simrel W^{\circ\prime}}
  \RuleTagHighlight{${\simrel}$subenv-val}
\end{gather*}
\caption{Simulation of configurations, heaps, and code terms.}
\label{fig:sim-code}
\end{figure}

The simulation relation between \LAlpha and \LCtx is defined in
Figures~\ref{fig:sim-code}, ~\ref{fig:sim-meta}, and~\ref{fig:sim-coerce}. The
highlighted rules extend the compilation rules, so that the simulation relation
is closed under reduction.

\newpage

\begin{figure}[tbp]
\footnotesize
\setlength{\abovedisplayskip}{2pt}\setlength{\belowdisplayskip}{2pt}%
\hfill\Fbox{$\Sigma; \Xi \vdash M : A \simrel N$}
\begin{gather*}
  \inference{k : \iota}
            {\Sigma; \Xi \vdash k : \iota \simrel k}
  \RuleTag{${\simrel}$const}
  \quad
  \inference{x{:}A \in \Xi}
            {\Sigma; \Xi \vdash x : A \simrel x}
  \RuleTag{${\simrel}$var}
  \\[1ex]
  \inference{\Xi \vdash A \compile A' & \Sigma; \Xi, x{:}A \vdash N : B \simrel N'}
            {\Sigma; \Xi \vdash \lambda (x{:}A).\,N : A \to B \simrel \lambda (x{:}A').\,N'}
  \RuleTag{${\simrel}$abs}
  \\[1ex]
  \inference{\Sigma; \Xi \vdash L : A \to B \simrel L' \\ \Sigma; \Xi \vdash M : A \simrel M'}
            {\Sigma; \Xi \vdash L\;M : B \simrel L'\;M'}
  \RuleTag{${\simrel}$app}
  \quad
  \inference{a{:}A \in \Sigma}
            {\Sigma; \Xi \vdash a : \REFTY{A} \simrel a}
  \RuleTagHighlight{${\simrel}$addr}
  \\[1ex]
  \inference{\Sigma; \Xi \vdash M : A \simrel M'}
            {\Sigma; \Xi \vdash \REF{M} : \REFTY{A} \simrel \REF{M'}}
  \RuleTag{${\simrel}$ref}
  \\[1ex]
  \inference{\Sigma; \Xi \vdash L : \REFTY{A} \simrel L' & \Sigma; \Xi \vdash M : A \simrel M'}
            {\Sigma; \Xi \vdash L := M : \Unit \simrel L' := M'}
  \RuleTag{${\simrel}$assign}
  \\[1ex]
  \inference{\Sigma; \Xi \vdash M : \REFTY{A} \simrel M'}
            {\Sigma; \Xi \vdash \mathop{!}M : A \simrel \mathop{!}M'}
  \RuleTag{${\simrel}$deref}
  \\[1ex]
  \inference{\Sigma; \Xi; e \vdash M^\circ : T \simrel M^{\circ\prime} &
             \Xi \vdash e \compile \Gamma^\circ}
            {\Sigma; \Xi \vdash \Quote{M^\circ}{e} : \Quote{T}{e} \simrel \CQuote{\Gamma^\circ}{M^{\circ\prime}}}
  \RuleTag{${\simrel}$quote}
  \\[1ex]
  \inference{\Sigma; \Xi, \ECalpha_{\mathsf{poly}} \vdash N : A \simrel N' & \ECalpha \notin \Xi}
            {\Sigma; \Xi \vdash \Lambda \ECalpha.\,N : \forall \ECalpha.\,A \simrel \Lambda \alpha.\,N'}
  \RuleTag{${\simrel}$ec-abs}
  \\[1ex]
  \inference{\Sigma; \Xi \vdash M : \forall \ECalpha.\,A \simrel M'
             & \Xi \vdash e \compile \Gamma^\circ}
            {\Sigma; \Xi \vdash M[e] : A[\ECalpha := e] \simrel M'[\Gamma^\circ]}
  \RuleTag{${\simrel}$ec-app}
  \\[1ex]
  \inference{\Xi \vdash e_1 \compile \Gamma^\circ_1 & \Xi \vdash e_2 \compile \Gamma^\circ_2 \\
             \Sigma; \Xi, (s : e_1 \sub e_2) \vdash N : A \simrel N' & s \notin \Xi}
            {\begin{array}[t]{@{}l@{}}
             \Sigma; \Xi \vdash (e_1 \sub e_2 \Rightarrow N) : (e_1 \sub e_2 \Rightarrow A) \simrel \lambda (s{:}\SubstTy{\Gamma^\circ_1}{\Gamma^\circ_2}).\,N'
             \end{array}}
  \RuleTag{${\simrel}$sub-intro}
  \\[1ex]
  \inference{\Sigma; \Xi \vdash M : (e_1 \sub e_2 \Rightarrow A) \simrel M' & \Xi \vdash e_1 \sub e_2 \compile N}
            {\Sigma; \Xi \vdash M\bullet : A \simrel M'\;N}
  \RuleTag{${\simrel}$sub-elim}
  \\[1ex]
  \inference{\Sigma; \Xi \vdash M : A \simrel M' & \Xi \vdash A \sub B \compile L}
            {\Sigma; \Xi \vdash M : B \simrel L\;M'}
  \RuleTag{${\simrel}$subsumption}
  \\[1ex]
  \inference{\Sigma; \Xi \vdash V : A \simrel V' & \Xi \vdash A \sub B \coerceval V' \Rightarrow W'}
            {\Sigma; \Xi \vdash V : B \simrel W'}
  \RuleTagHighlight{${\simrel}$subsumption-val}
\end{gather*}
\caption{Simulation of meta terms.}
\label{fig:sim-meta}
\end{figure}

\begin{figure}[tbp]
\footnotesize
\setlength{\abovedisplayskip}{2pt}\setlength{\belowdisplayskip}{2pt}%
\hfill\Fbox{$\Xi \vdash A \sub B \coerceval V \Rightarrow W$}
\vspace{10pt}
\begin{gather*}
  \inference{}
            {\Xi \vdash \iota \sub \iota \coerceval V \Rightarrow V}
  \RuleTag{${\coerceval}$base}
  \\[2ex]
  \inference{\Xi \vdash A \sub B \compile M & \Xi \vdash B \sub A \compile N}
            {\Xi \vdash \REFTY{A} \sub \REFTY{B} \coerceval V \Rightarrow \Proxy{V}{M}{N}}
  \RuleTag{${\coerceval}$ref}
  \\[2ex]
  \inference{\Xi \vdash e_1 \compile \Gamma^\circ & \Xi \vdash e_2 \compile \Delta^\circ
             & \Xi \vdash e_1 \sub e_2 \compile V}
            {\Xi \vdash \Quote{T}{e_1} \sub \Quote{T}{e_2} \coerceval
             \CQuote{\Gamma^\circ}{V^\circ} \Rightarrow \CQuote{\Delta^\circ}{\Sem{V}(V^\circ)}}
  \RuleTag{${\coerceval}$code}
  \\[2ex]
  \inference{\Xi \vdash B_1 \compile B_1' & \Xi \vdash B_1 \sub A_1 \compile L & \Xi \vdash A_2 \sub B_2 \compile M}
            {
             \Xi \vdash A_1 \to A_2 \sub B_1 \to B_2 \coerceval
             V \Rightarrow \lambda (x{:}B_1').\,M\;(V\;(L\;x))
             }
  \RuleTag{${\coerceval}$fun}
  \\[2ex]
  \inference{\Xi, \ECalpha_{\mathsf{poly}} \vdash A \sub B \compile M & \ECalpha \notin \Xi}
            {\Xi \vdash \forall \ECalpha.\,A \sub \forall \ECalpha.\,B \coerceval
             V \Rightarrow \Lambda \alpha.\,M\;(V[\alpha])}
  \RuleTag{${\coerceval}$forall}
  \\[2ex]
  \inference{\Xi \vdash e_1 \sub e_2 \compile C
             & \Xi, (s : e_1 \sub e_2) \vdash A \sub B \compile M & s \notin \Xi}
            {\Xi \vdash (e_1 \sub e_2 \Rightarrow A) \sub (e_1 \sub e_2 \Rightarrow B) \coerceval
             V \Rightarrow \lambda (s{:}C).\,M\;(V\;s)}
  \RuleTag{${\coerceval}$constraint}
\end{gather*}
\caption{The coerced value relation.}
\label{fig:sim-coerce}
\end{figure}

\end{document}